\documentclass[
 aps,
 11pt,
 final,
 notitlepage,
 oneside,
 twocolumn,
 nobibnotes,
 nofootinbib,
 superscriptaddress,
 noshowpacs,
 centertags]
{revtex4}

\def\squareforqed{\hbox{\rlap{$\sqcap$}$\sqcup$}}

\def\sq{\ifmmode\squareforqed\else{\unskip\nobreak\hfil
\penalty50\hskip1em\null\nobreak\hfil\squareforqed
\parfillskip=0pt\finalhyphendemerits=0\endgraf}\fi}

\def\degr{\hbox{$^\circ$}}

\def\arcsec{\hbox{$^{\prime\prime}$}}

\def\utw{\smash{\rlap{\lower5pt\hbox{$\sim$}}}}

\def\udtw{\smash{\rlap{\lower6pt\hbox{$\approx$}}}}

\def\fd{\hbox{$\,.\!\!^{\rm d}$}}

\def\farcs{\hbox{$\,.\!\!^{\prime\prime}$}}

\def\diameter{{\ifmmode\mathchoice
{\ooalign{\hfil\hbox{$\displaystyle/$}\hfil\crcr
{\hbox{$\displaystyle\mathchar"20D$}}}}
{\ooalign{\hfil\hbox{$\textstyle/$}\hfil\crcr
{\hbox{$\textstyle\mathchar"20D$}}}}
{\ooalign{\hfil\hbox{$\scriptstyle/$}\hfil\crcr
{\hbox{$\scriptstyle\mathchar"20D$}}}}
{\ooalign{\hfil\hbox{$\scriptscriptstyle/$}\hfil\crcr
{\hbox{$\scriptscriptstyle\mathchar"20D$}}}}
\else{\ooalign{\hfil/\hfil\crcr\mathhexbox20D}}%
\fi}}

\newcommand{\ab}{Astrophysical Bulletin }

\newcommand{\aaa}{Astron. and Astrophys. }

\newcommand{\aas}{Astron. and Astrophys. Suppl. }

\newcommand{\apjs}{Astrophys.~J. Suppl. }

\newcommand{\mnras}{Monthly Notices Royal Astron. Soc. }

\newcommand{\alet}{Astronomy Letters }
\newcommand{\an}{Astronomische Nachrichten }

\begin{document}
\selectlanguage{english}


\title{Results of Magnetic Field Measurements Performed with the 6-m Telescope.\\
IV. Observations in 2010}

\author{\firstname{I.~I.}~\surname{Romanyuk}}
\email{roman@sao.ru} \affiliation{Special Astrophysical Observatory,
Russian Academy of Sciences,
              Nizhnii Arkhyz, 369167 Russia}

\author{\firstname{E.~A.}~\surname{Semenko}}
\affiliation{Special Astrophysical Observatory, Russian Academy of
Sciences,
              Nizhnii Arkhyz, 369167 Russia}

\author{\firstname{D.~O.}~\surname{Kudryavtsev}}
\affiliation{Special Astrophysical Observatory, Russian Academy of
Sciences,
              Nizhnii Arkhyz, 369167 Russia}

\author{\firstname{A.~V.}~\surname{Moiseeva}}
\affiliation{Special Astrophysical Observatory, Russian Academy of
Sciences,
              Nizhnii Arkhyz, 369167 Russia}

\author{\firstname{I.~A.}~\surname{Yakunin}}
\affiliation{Special Astrophysical Observatory, Russian Academy of
Sciences,
              Nizhnii Arkhyz, 369167 Russia}

\begin{abstract}
We present the results of measurements of magnetic fields, radial
velocities and rotation velocities  for 92 objects, mainly
main-sequence chemically peculiar stars. Observations were performed
at  the 6-m BTA telescope using Main Stellar Spectrograph with a
Zeeman analyzer. In 2010, twelve new magnetic stars were discovered:
HD\,17330, HD\,29762, HD\,49884, HD\,54824, HD\,89069, HD\,96003,
HD\,113894, HD\,118054, HD\,135679, HD\,138633, HD\,138777,
BD\,+53.1183. The presence of a field is suspected in HD\,16705,
HD\,35379 and HD\,35881. Observations of standard stars without a
magnetic field confirm the absence of systematic errors which can
introduce distortions into the measurements of  longitudinal field.
The paper gives comments on the results of investigation of each
star.
\end{abstract}

\maketitle

\section{INTRODUCTION}

We continue an ensemble of papers on the study of magnetic stars at
the 6-m telescope, initiated in Romanyuk et al.~\cite{1:Romanyuk_n_en_4,2:Romanyuk_n_en_4,3:Romanyuk_n_en_4} by a
publication of complete reports on the measurement of magnetic fields
from the 2007--2009 observational data. This paper presents the
results of measurements of  magnetic fields in chemically peculiar
stars, observed with the Main Stellar Spectrograph~(MSS) of the 6-m
BTA telescope of the Special Astrophysical Observatory of the Russian
Academy of Sciences (SAO RAS)  in 2010. The scientific substantiation
of studies of our series is given in previous publications.
Equipment, methods of observations and data reduction in general
terms have not been changed in comparison with previous years.
However, some changes did take place. Firstly, since March 2010 a new
large-format CCD sized \mbox{$4600\times2000$}~pixels is used
 as a detector at the MSS.
Secondly, the data reduction technique, described in
\cite{1:Romanyuk_n_en_4,2:Romanyuk_n_en_4,3:Romanyuk_n_en_4}  has been generally saved, however, a part of the field measurements
 was also performed using the
regression method~\cite{4:Romanyuk_n_en_4}.

\section{OBSERVATIONS AND REDUCTION METHOD}

The material that served as the basis for this study was obtained at
the 6-m telescope in 2010 during 21 nights of observations within
four main programs: Magnetic Fields of Massive Stars (principal
applicant: I.~I.~Romanyuk, SAO~RAS); New Magnetic Stars (principal
applicant: D.~O.~Kudryavtsev, SAO~RAS); Selected Magnetic Stars
(principal applicant: E.~A.~Semenko, SAO~RAS); Geometry of Magnetic
Fields in CP Stars (principal applicant: G.~Wade, Canada). One
hundred eighty-four pairs of circularly polarized spectra were
obtained for 92 stars. The list of objects is significantly different
from the similar set of previous years.

A part of  observations (in January and February 2010) was conducted,
just like in 2009, using a CCD sized $2000\times2000$~pixels. Since
March 25, a new CCD array sized $4600\times2000$~pixels has been used
for the observations. An introduction of a new panoramic CCD allowed
to expand the simultaneously registered spectral range up to 550~\AA,
which significantly improved the possibilities for observing magnetic
fields of  stars. In most cases, we worked in the spectral range from
4450~\AA\ to 5000~\AA.

We have earlier repeatedly described the preliminary reduction
procedure of the spectral data (image  processing, extraction of
spectra, wavelength calibration, etc.) and measurement of magnetic
fields by the modified Babcock method, for example, in~\cite{1:Romanyuk_n_en_4,2:Romanyuk_n_en_4,3:Romanyuk_n_en_4}.
In addition, new software that allows determining the longitudinal
magnetic field of stars by the regression method proposed by Bagnulo
et al.~\cite{4:Romanyuk_n_en_4} was applied. This method proved to be very useful in
the studies of fast rotators with complex line profiles for which the
classical
  measurement procedure led to a large scattering of  results.

When measuring the magnetic fields of stars with narrow lines, both methods
give approximately the same results. However,
there are significant differences for fast rotators. As a rule, the
longitudinal fields, obtained by the regression method, turn out to be
substantially smaller than those measured by the classical method.
In addition to the longitudinal magnetic fields $B_e$, we determined
 radial velocities $V_r$ and  projections of  rotation velocities onto
the line of sight $v_e \sin i$  for each star.

\section{RESULTS OF MEASUREMENTS}

The results of magnetic field measurements are presented in the summary
Table. The columns of the table contain data on the names of  stars in the
order of increasing numbers in the  HD and BD catalogs, on the Julian date
of observations, the longitudinal field $B_e$, the radial velocity $V_r$, and
rotation velocity $v_e \sin i$ with measurement errors $\sigma$, and
the $S/N$ ratio of the resulting spectrum. The longitudinal field,
measured by the standard Babcock method  is marked with the letter ``z'',
by the regression method---by the letter ``r'', measurements of the magnetic
field by
the core of the hydrogen  H$\beta$ line---by the letter ``h''.
Non-magnetic standard
stars  in the table are marked with the symbol ``*'', standard
stars with a well-known   variability law of the longitudinal
field component are denoted by
``**''.  The stars  for which the magnetic field
was discovered for the first time are printed in italics.

 \onecolumngrid
\setcaptionwidth{\linewidth}%
\setcaptionmargin{0mm} %
\onelinecaptionsfalse \captionstyle{nonumber}
\medskip
\begin{longtable*}{l|c|r@{$\,\pm\,$}l|r@{$\,\pm\,$}l|r@{$\,\pm\,$}l|c}
\caption{\centerline{The results of measurements of magnetic fields, radial velocities,} \centerline{and rotation velocities of stars according to observations 2010}}\label{tab1:Romanyuk_n_en_4}\\
\hline
\multicolumn{1}{c|}{Star} & JD~(2450000+) & \multicolumn{2}{c|}{$B_e \pm \sigma$,} & \multicolumn{2}{c|}{$V_r \pm \sigma$,} & \multicolumn{2}{c|}{$v_e\sin i \pm \sigma$,} & $S/N$ \\
                                             &                                & \multicolumn{2}{c|}{G}                & \multicolumn{2}{c|}{~km\,s$^{-1}$}     & \multicolumn{2}{c|}{km\,s$^{-1}$}            &  \\
 \hline
\multicolumn{1}{c|}{(1)} & (2) & \multicolumn{2}{c|}{(3)} & \multicolumn{2}{c|}{(4)} & \multicolumn{2}{c|}{(5)} & (6) \\
\endfirsthead
\caption{\centerline{Contd.}}\\
\hline
\multicolumn{1}{c|}{Star} & JD~(2450000+) & \multicolumn{2}{c|}{$B_e \pm \sigma$,} & \multicolumn{2}{c|}{$V_r \pm \sigma$,} & \multicolumn{2}{c|}{$v_e\sin i \pm \sigma$,} & $S/N$ \\
                                             &                                & \multicolumn{2}{c|}{G}                & \multicolumn{2}{c|}{~km\,s$^{-1}$}     & \multicolumn{2}{c|}{km\,s$^{-1}$}            &  \\
\hline
\multicolumn{1}{c|}{(1)} & (2) & \multicolumn{2}{c|}{(3)} & \multicolumn{2}{c|}{(4)} & \multicolumn{2}{c|}{(5)} & (6) \\
\hline
\endhead
\hline
\endfoot
\endlastfoot
\hline

HD\,653                              & 5488.445  &   $-30$ & $110$  (z)   & $+30.2$  & $2.6$      & $80 $ & $ 13$               & 200  \\
                                     &           &  $-100$ & $140$  (r)   & \multicolumn{2}{c|}{} & \multicolumn{2}{c|}{}       &      \\
                                     & 5554.220  &  $+130$ & $110$  (z)   & $-17.0$  & $3.4$      & $70 $ & $ 10$               & 220  \\
                                     &           &  $+320$ & $80$   (r)   & \multicolumn{2}{c|}{} & \multicolumn{2}{c|}{}       &      \\
HD\,965                              & 5431.458  &   $-70$ & $20$   (z)   & $-1.0$   & $1.8$      & \multicolumn{2}{c|}{$< 20$} & 170  \\
                                     & 5459.492  &  $-240$ & $50$   (z)   & $-4.5$   & $2.6$      & \multicolumn{2}{c|}{}       & 200  \\
                                     & 5461.462  &  $-140$ & $50$   (z)   & $-3.4$   & $2.9$      & \multicolumn{2}{c|}{}       & 230  \\
                                     & 5553.174  &  $-340$ & $20$   (z)   & $-1.3$   & $1.0$      & \multicolumn{2}{c|}{}       & 200  \\
                                     &           &  $-330$ & $40$   (r)   & \multicolumn{2}{c|}{} & \multicolumn{2}{c|}{}       &      \\
                                     & 5555.139  &  $-360$ & $30$   (z)   & $-3.2$   & $2.4$      & \multicolumn{2}{c|}{}       & 140  \\
                                     &           &  $-330$ & $20$   (r)   & \multicolumn{2}{c|}{} & \multicolumn{2}{c|}{}       &      \\
HD\,5441                             & 5488.479  &  $-440$ & $20$   (z)   & $+43.6$  & $2.4$      & $21 $ & $ 3$                & 190  \\
                                     &           &  $-450$ & $20$   (r)   & \multicolumn{2}{c|}{} & \multicolumn{2}{c|}{}       &      \\
HD\,5797                             & 5253.628  &  \multicolumn{2}{c|}{} & $-3.3$   & $1.1$      & \multicolumn{2}{c|}{$< 20$} & 220  \\
                                     & 5255.218  &  \multicolumn{2}{c|}{} & \multicolumn{2}{c|}{} & \multicolumn{2}{c|}{}       &      \\
HD\,6757                             & 5431.550  & $+2800$ & $90$   (z)   & $-9.4$   & $2.4$      & $28 $ & $ 5$                & 330  \\
HD\,16705                            & 5553.192  & $-3720$ & $1660$ (z)   & $-12.2$  & $2.9$      & $100 $ & $ 8$               & 350  \\
                                     &           &  $+720$ & $200$  (r)   & \multicolumn{2}{c|}{} & \multicolumn{2}{c|}{}       &      \\
\it{HD\,17330}                       & 5553.225  &  $+150$ & $320$  (z)   & $-13.6 $ & $ 2.8$     & \multicolumn{2}{c|}{$< 20$} & 400  \\
                                     &           &  $-420$ & $30$   (r)   & \multicolumn{2}{c|}{} & \multicolumn{2}{c|}{}       &      \\
HD\,23924                            & 5554.435  &   $-60$ & $50$   (z)   & $-2.1  $ & $ 2.0$     & $40 $ & $ 6$                & 300  \\
                                     &           &   $-50$ & $60$   (r)   & \multicolumn{2}{c|}{} & \multicolumn{2}{c|}{}       &      \\
HD\,23964                            & 5555.475  &   $-10$ & $30$   (z)   & $+10.2 $ & $ 1.9$     & $20 $ & $ 3$                & 250  \\
\it{HD\,29762}                       & 5555.456  &  $+300$ & $50$   (z)   & $-9.2  $ & $ 1.7$     & $32 $ & $ 4$                & 250  \\
                                     &           &  $+190$ & $60$   (r)   & \multicolumn{2}{c|}{} & \multicolumn{2}{c|}{}       &      \\
HD\,32549                            & 5488.619  &  $+140$ & $80$   (z)   & $+24.3 $ & $ 2.8$     & $61 $ & $ 6$                & 600  \\
                                     &           &   $+60$ & $70$   (r)   & \multicolumn{2}{c|}{} & \multicolumn{2}{c|}{}       &      \\
                                     & 5554.495  &  $-250$ & $120$  (z)   & $+10.0 $ & $ 2.9$     & \multicolumn{2}{c|}{}       & 800  \\
                                     &           &  $+280$ & $150$  (r)   & \multicolumn{2}{c|}{} & \multicolumn{2}{c|}{}       &      \\
HD\,33256*                           & 5202.267  &   $+20$ & $20$   (z)   & $-21.2 $ & $ 2.4$     & \multicolumn{2}{c|}{$< 20$} & 500  \\
                                     &           &  $+110$ & $160$  (h)   & \multicolumn{2}{c|}{} & \multicolumn{2}{c|}{}       &      \\
                                     & 5488.571  &   $-10$ & $10$   (z)   & $+14.6 $ & $ 2.5$     & \multicolumn{2}{c|}{}       & 400  \\
                                     &           &   $-10$ & $10$   (r)   & \multicolumn{2}{c|}{} & \multicolumn{2}{c|}{}       &      \\
                                     & 5554.253  &   $-50$ & $10$   (z)   & $+3.6  $ & $ 1.5$     & \multicolumn{2}{c|}{}       & 500  \\
                                     &           &   $-50$ & $10$   (r)   & \multicolumn{2}{c|}{} & \multicolumn{2}{c|}{}       &      \\
                                     & 5555.245  &     $0$ & $10$   (z)   & $+6.0  $ & $ 3.1$     & \multicolumn{2}{c|}{}       & 800  \\
                                     &           &   $+10$ & $10$   (r)   & \multicolumn{2}{c|}{} & \multicolumn{2}{c|}{}       &      \\
HD\,34307                            & 5553.313  &   $-50$ & $90$   (z)   & $+29.8 $ & $ 2.8$     & $21 $ & $ 4$                & 300  \\
                                     &           &  $+160$ & $180$  (r)   & \multicolumn{2}{c|}{} & \multicolumn{2}{c|}{}       &      \\
HD\,34968                            & 5553.395  &   $+30$ & $250$  (z)   & $+27.9 $ & $ 3.8$     & $105 $ & $ 20$              & 300  \\
                                     &           &  $-170$ & $240$  (r)   & \multicolumn{2}{c|}{} & \multicolumn{2}{c|}{}       &      \\
HD\,35101                            & 5553.577  &  $+600$ & $950$  (z)   & $-18.9 $ & $ 2.8$     & $110 $ & $ 30$              & 400  \\
                                     &           &   $-30$ & $190$  (r)   & \multicolumn{2}{c|}{} & \multicolumn{2}{c|}{}       &      \\
                                     & 5554.592  &  $+740$ & $530$  (z)   & $-25.7 $ & $ 2.5$     & \multicolumn{2}{c|}{}       & 400  \\
                                     &           &  $-430$ & $190$  (r)   & \multicolumn{2}{c|}{} & \multicolumn{2}{c|}{}       &      \\
HD\,35298                            & 5554.300  & $-6090$ & $300$  (z)   & $+20.9 $ & $ 3.4$     & $50 $ & $ 7$                & 400  \\
                                     &           & $-3440$ & $150$  (r)   & \multicolumn{2}{c|}{} & \multicolumn{2}{c|}{}       &      \\
HD\,35379                            & 5552.565  &  $-200$ & $110$  (z)   & $+3.7  $ & $ 2.8$     & $45 $ & $ 5$                & 300  \\
                                     &           &  $-250$ & $120$  (r)   & \multicolumn{2}{c|}{} & \multicolumn{2}{c|}{}       &      \\
HD\,35456                            & 5554.338  &  $+650$ & $70$   (z)   & $+11.0 $ & $ 2.8$     & $22 $ & $ 2$                & 400  \\
                                     &           &  $+640$ & $80$   (r)   & \multicolumn{2}{c|}{} & \multicolumn{2}{c|}{}       &      \\
HD\,35548                            & 5553.294  &   $-10$ & $20$   (z)   & $-9.3  $ & $ 2.4$     & \multicolumn{2}{c|}{$< 20$} & 500  \\
                                     &           &   $+20$ & $60$   (r)   & \multicolumn{2}{c|}{} & \multicolumn{2}{c|}{}       &      \\
HD\,35575                            & 5553.304  &  $-200$ & $490$  (z)   & $+24.2 $ & $ 2.9$     & $150 $ & $ 10$              & 500  \\
                                     &           &  $+310$ & $270$  (r)   & \multicolumn{2}{c|}{} & \multicolumn{2}{c|}{}       &      \\
HD\,35730                            & 5553.238  &  $+150$ & $320$  (z)   & $+22.7 $ & $ 2.7$     & $54 $ & $ 6$                & 400  \\
                                     &           &   $-30$ & $220$  (r)   & \multicolumn{2}{c|}{} & \multicolumn{2}{c|}{}       &      \\
HD\,35881                            & 5553.258  & $-1070$ & $590$  (z)   & $+19.5 $ & $ 3.7$     & $205 $ & $ 20$              & 500  \\
                                     &           & $-1130$ & $370$  (r)   & \multicolumn{2}{c|}{} & \multicolumn{2}{c|}{}       &      \\
HD\,36032                            & 5553.454  &  $-900$ & $200$  (z)   & $+29.6 $ & $ 3.4$     & $205 $ & $ 16$              & 300  \\
                                     &           &  $-110$ & $230$  (r)   & \multicolumn{2}{c|}{} & \multicolumn{2}{c|}{}       &      \\
HD\,36313                            & 5554.320  &  $+120$ & $120$  (z)   & $+43.0 $ & $ 2.9$     & $27 $ & $ 2$                & 400  \\
                                     &           &  $+560$ & $180$  (r)   & \multicolumn{2}{c|}{} & \multicolumn{2}{c|}{}       &      \\
HD\,36485                            & 5553.247  & $-2350$ & $250$  (z)   & $+22.1 $ & $ 3.5$     & $40 $ & $ 3$                & 400  \\
                                     &           & $-2310$ & $120$  (r)   & \multicolumn{2}{c|}{} & \multicolumn{2}{c|}{}       &      \\
                                     & 5553.480  & $-2330$ & $220$  (z)   & \multicolumn{2}{c|}{} & \multicolumn{2}{c|}{}       & 400  \\
                                     &           & $-2210$ & $190$  (r)   & \multicolumn{2}{c|}{} & \multicolumn{2}{c|}{}       &      \\
                                     & 5554.263  & $-2400$ & $210$  (z)   & $+15.7 $ & $ 2.6$     & \multicolumn{2}{c|}{}       & 400  \\
                                     &           & $-2270$ & $120$  (r)   & \multicolumn{2}{c|}{} & \multicolumn{2}{c|}{}       &      \\
                                     & 5554.481  & $-2700$ & $230$  (z)   & \multicolumn{2}{c|}{} & \multicolumn{2}{c|}{}       & 400  \\
                                     &           & $-2570$ & $180$  (r)   & \multicolumn{2}{c|}{} & \multicolumn{2}{c|}{}       &      \\
                                     & 5555.254  & $-2830$ & $260$  (z)   & $+16.4 $ & $ 2.4$     & \multicolumn{2}{c|}{}       & 400  \\
                                     &           & $-2470$ & $160$  (r)   & \multicolumn{2}{c|}{} & \multicolumn{2}{c|}{}       &      \\
                                     & 5555.485  & $-2830$ & $320$  (z)   & \multicolumn{2}{c|}{} & \multicolumn{2}{c|}{}       & 300  \\
                                     &           & $-2370$ & $120$  (r)   & \multicolumn{2}{c|}{} & \multicolumn{2}{c|}{}       &      \\
HD\,36526                            & 5553.338  & $+2730$ & $320$  (z)   & $+21.5 $ & $ 3.4$     & $45 $ & $ 5$                & 400  \\
                                     &           & $+2180$ & $170$  (r)   & \multicolumn{2}{c|}{} & \multicolumn{2}{c|}{}       &      \\
HD\,36540                            & 5553.367  &  $+400$ & $250$  (z)   & $+14.0 $ & $ 2.9$     & $80 $ & $ 15$               & 400  \\
                                     &           &  $+650$ & $300$  (r)   & \multicolumn{2}{c|}{} & \multicolumn{2}{c|}{}       &      \\
HD\,36629                            & 5553.385  &   $+80$ & $50$   (z)   & $+35.0 $ & $ 3.7$     & $23 $ & $ 5$                & 400  \\
                                     &           &   $+70$ & $100$  (r)   & \multicolumn{2}{c|}{} & \multicolumn{2}{c|}{}       &      \\
HD\,36916                            & 5554.347  &  $-950$ & $150$  (z)   & $+12.7 $ & $ 2.9$     & $42 $ & $ 5$                & 400  \\
                                     &           &  $-660$ & $220$  (r)   & \multicolumn{2}{c|}{} & \multicolumn{2}{c|}{}       &      \\
HD\,36982                            & 5554.368  &  $+170$ & $330$  (z)   & $+12.6 $ & $ 3.4$     & $160 $ & $ 20$              & 300  \\
                                     &           &  $+200$ & $100$  (r)   & \multicolumn{2}{c|}{} & \multicolumn{2}{c|}{}       &      \\
HD\,37022                            & 5282.217  &  $-780$ & $270$  (z)   & $+29.6 $ & $ 2.8$     & $0 $ & $ 10$                & 1000 \\
                                     &           &  $-510$ & $110$  (r)   & \multicolumn{2}{c|}{} & \multicolumn{2}{c|}{}       &      \\
                                     & 5284.253  &  $-250$ & $280$  (z)   & $+25.7 $ & $ 2.7$     & \multicolumn{2}{c|}{}       & 800  \\
                                     &           &   $-40$ & $150$  (r)   & \multicolumn{2}{c|}{} & \multicolumn{2}{c|}{}       &      \\
                                     & 5552.460  &  $-190$ & $250$  (z)   & $+19.1 $ & $ 3.7$     & \multicolumn{2}{c|}{}       & 1000 \\
                                     &           &   $+20$ & $230$  (r)   & \multicolumn{2}{c|}{} & \multicolumn{2}{c|}{}       &      \\
                                     & 5553.279  &  $+560$ & $560$  (z)   & $+20.7 $ & $ 2.8$     & \multicolumn{2}{c|}{}       & 1200 \\
                                     &           &  $+110$ & $170$  (r)   & \multicolumn{2}{c|}{} & \multicolumn{2}{c|}{}       &      \\
                                     & 5554.275  &   $+90$ & $260$  (z)   & $+12.4 $ & $ 3.7$     & \multicolumn{2}{c|}{}       & 1000 \\
                                     &           &   $-10$ & $180$  (r)   & \multicolumn{2}{c|}{} & \multicolumn{2}{c|}{}       &      \\
                                     & 5555.270  &  $+630$ & $170$  (z)   & $+16.2 $ & $ 3.7$     & \multicolumn{2}{c|}{}       & 1000 \\
                                     &           &   $-80$ & $180$  (r)   & \multicolumn{2}{c|}{} & \multicolumn{2}{c|}{}       &      \\
HD\,37140                            & 5555.298  &  $-580$ & $90$   (z)   & $+23.4 $ & $ 2.4$     & $25 $ & $ 2$                & 400  \\
                                     &           &  $-350$ & $90$   (r)   & \multicolumn{2}{c|}{} & \multicolumn{2}{c|}{}       &      \\
HD\,37151                            & 5555.503  &     $0$ & $80$   (z)   & $+18.2 $ & $ 2.7$     & $30 $ & $ 3$                & 400  \\
                                     &           &   $+30$ & $50$   (r)   & \multicolumn{2}{c|}{} & \multicolumn{2}{c|}{}       &      \\
HD\,37479                            & 5555.325  & $-1050$ & $1080$ (z)   & \multicolumn{2}{c|}{} & \multicolumn{2}{c|}{}       &      \\
                                     &           &  $+140$ & $330$  (r)   & $+15.0 $ & $ 2.9$     & $100 $ & $ 25$              & 400  \\
HD\,37525                            & 5555.342  &  $+670$ & $1670$ (z)   & $+39.0 $ & $ 3.4$     & $150 $ & $ 30$              & 400  \\
                                     &           &   $+20$ & $290$  (r)   & \multicolumn{2}{c|}{} & \multicolumn{2}{c|}{}       &      \\
HD\,37687                            & 5555.507  &  $+580$ & $40$   (z)   & $+18.1 $ & $ 2.3$     & $22 $ & $ 3$                & 400  \\
                                     &           &  $+490$ & $40$   (r)   & \multicolumn{2}{c|}{} & \multicolumn{2}{c|}{}       &      \\
HD\,37776                            & 5282.183  & $+8600$ & $1700$ (z)   & $+26.5 $ & $ 3.1$     & \multicolumn{2}{c|}{}       & 600  \\
                                     &           & $+2090$ & $170$  (r)   & \multicolumn{2}{c|}{} & \multicolumn{2}{c|}{}       &      \\
                                     & 5284.233  & $-6200$ & $7100$ (z)   & $+23.6 $ & $ 3.7$     & \multicolumn{2}{c|}{}       & 500  \\
                                     &           &  $-130$ & $290$  (r)   & \multicolumn{2}{c|}{} & \multicolumn{2}{c|}{}       &      \\
HD\,38104                            & 5202.600  &  $+160$ & $110$  (z)   & $+3.8  $ & $ 1.1$     & $35 $ & $ 5$                & 700  \\
                                     &           &  $-110$ & $210$  (h)   & \multicolumn{2}{c|}{} & \multicolumn{2}{c|}{}       &      \\
                                     & 5288.306  &     $0$ & $50$   (z)   & $-2.4  $ & $ 1.4$     & \multicolumn{2}{c|}{}       & 500  \\
                                     &           &  $+100$ & $280$  (h)   & \multicolumn{2}{c|}{} & \multicolumn{2}{c|}{}       &      \\
HD\,38823                            & 5202.600  &  $-220$ & $80$   (z)   & $+3.6  $ & $ 1.2$     & $36 $ & $ 3$                & 300  \\
                                     &           &  $-280$ & $150$  (h)   & \multicolumn{2}{c|}{} & \multicolumn{2}{c|}{}       &      \\
HD\,39317                            & 5554.500  &   $+10$ & $110$  (z)   & $-12.1 $ & $ 2.6$     & $70 $ & $ 15$               & 500  \\
                                     &           &   $+50$ & $160$  (r)   & \multicolumn{2}{c|}{} & \multicolumn{2}{c|}{}       &      \\
HD\,45583                            & 5202.348  & $-2400$ & $200$  (z)   & $+21.4 $ & $ 2.8$     & $80 $ & $ 10$               & 400  \\
                                     &           & $-1650$ & $260$  (h)   & \multicolumn{2}{c|}{} & \multicolumn{2}{c|}{}       &      \\
                                     & 5284.272  & $+5320$ & $470$  (z)   & $+24.8 $ & $ 3.8$     & \multicolumn{2}{c|}{}       & 300  \\
                                     &           & $+2700$ & $100$  (r)   & \multicolumn{2}{c|}{} & \multicolumn{2}{c|}{}       &      \\
                                     & 5488.592  & $+5710$ & $620$  (z)   & $+32.3 $ & $ 2.7$     & \multicolumn{2}{c|}{}       & 500  \\
                                     &           & $+2160$ & $170$  (r)   & \multicolumn{2}{c|}{} & \multicolumn{2}{c|}{}       &      \\
                                     & 5552.485  & $+2770$ & $290$  (z)   & $+23.4 $ & $ 2.8$     & \multicolumn{2}{c|}{}       & 400  \\
                                     &           & $+2650$ & $190$  (r)   & \multicolumn{2}{c|}{} & \multicolumn{2}{c|}{}       &      \\
                                     & 5553.425  & $+4807$ & $403$  (z)   & $+30.0 $ & $ 2.9$     & \multicolumn{2}{c|}{}       & 500  \\
                                     &           & $+2670$ & $190$  (r)   & \multicolumn{2}{c|}{} & \multicolumn{2}{c|}{}       &      \\
\it{HD\,49884}                       & 5283.302  &  $-100$ & $50$   (z)   & $-6.9  $ & $ 2.7$     & \multicolumn{2}{c|}{$< 20$} & 300  \\
                                     &           &  $-180$ & $30$   (r)   & \multicolumn{2}{c|}{} & \multicolumn{2}{c|}{}       &      \\
                                     & 5284.297  &  $-190$ & $40$   (z)   & $+0.3  $ & $ 0.1$     & \multicolumn{2}{c|}{}       & 400  \\
                                     &           &  $-125$ & $10$   (r)   & \multicolumn{2}{c|}{} & \multicolumn{2}{c|}{}       &      \\
                                     & 5552.625  &  $-310$ & $40$   (z)   & $-5.5  $ & $ 2.7$     & \multicolumn{2}{c|}{}       & 200  \\
                                     &           &  $-290$ & $30$   (r)   & \multicolumn{2}{c|}{} & \multicolumn{2}{c|}{}       &      \\
HD\,50169                            & 5282.238  &  $+130$ & $50$   (z)   & $+14.6 $ & $ 2.4$     & \multicolumn{2}{c|}{$< 20$} & 300  \\
                                     &           &   $+80$ & $15$   (r)   & \multicolumn{2}{c|}{} & \multicolumn{2}{c|}{}       &      \\
HD\,50461                            & 5554.418  & $+1550$ & $770$  (z)   & $+29.5 $ & $ 3.4$     & $80 $ & $ 15$               & 400  \\
                                     &           &  $+290$ & $220$  (r)   & \multicolumn{2}{c|}{} & \multicolumn{2}{c|}{}       &      \\
HD\,51418                            & 5281.438  &  $+520$ & $50$   (z)   & $-28.3 $ & $ 2.6$     & $28 $ & $ 5$                & 300  \\
                                     & 5282.306  &  $-460$ & $70$   (z)   & $-33.2 $ & $ 2.7$     & \multicolumn{2}{c|}{}       & 700  \\
                                     &           &  $-520$ & $40$   (r)   & \multicolumn{2}{c|}{} & \multicolumn{2}{c|}{}       &      \\
                                     & 5283.318  &  $+360$ & $60$   (z)   & $-30.7 $ & $ 3.1$     & \multicolumn{2}{c|}{}       & 400  \\
                                     &           &  $+320$ & $30$   (r)   & \multicolumn{2}{c|}{} & \multicolumn{2}{c|}{}       &      \\
                                     & 5284.354  &  $-250$ & $40$   (z)   & $-25.1 $ & $ 2.4$     & \multicolumn{2}{c|}{}       & 500  \\
                                     &           &  $-200$ & $20$   (r)   & \multicolumn{2}{c|}{} & \multicolumn{2}{c|}{}       &      \\
HD\,52711                            & 5288.294  &   $-40$ & $20$   (z)   & $+30.9 $ & $ 2.7$     & $33 $ & $ 5$                & 400  \\
\it{HD\,54824}                       & 5283.279  &  $-690$ & $180$  (z)   & $+27.0 $ & $ 2.8$     & $50 $ & $ 10$               & 300  \\
                                     &           &  $-482$ & $50$   (r)   & \multicolumn{2}{c|}{} & \multicolumn{2}{c|}{}       &      \\
                                     & 5552.604  &  $+470$ & $120$  (z)   & $+29.7 $ & $ 2.9$     & \multicolumn{2}{c|}{}       & 350  \\
                                     &           &  $+200$ & $160$  (r)   & \multicolumn{2}{c|}{} & \multicolumn{2}{c|}{}       &      \\
                                     & 5553.598  &  $-630$ & $70$   (z)   & $+35.5 $ & $ 2.9$     & \multicolumn{2}{c|}{}       & 300  \\
                                     &           &  $-470$ & $110$  (r)   & \multicolumn{2}{c|}{} & \multicolumn{2}{c|}{}       &      \\
HD\,62512                            & 5553.545  &   $-80$ & $50$   (z)   & $+12.0 $ & $ 2.8$     & $22$ & $ 5$                 & 300  \\
                                     &           &  $-100$ & $50$   (r)   & \multicolumn{2}{c|}{} &  \multicolumn{2}{c|}{}      &      \\
                                     & 5554.518  &  $-180$ & $40$   (z)   & $+4.2  $ & $ 3.6$     &  \multicolumn{2}{c|}{}      & 300  \\
                                     &           &  $-200$ & $70$   (r)   & \multicolumn{2}{c|}{} &  \multicolumn{2}{c|}{}      &      \\
HD\,65339**                          & 5202.608  & $+3740$ & $100$  (z)   & $-1.2  $ & $ 1.0$     & $25$ & $ 5$                 & 600  \\
                                     &           & $+2740$ & $300$  (h)   & \multicolumn{2}{c|}{} & \multicolumn{2}{c|}{}       &      \\
                                     & 5281.451  & $+2340$ & $190$  (z)   & $-2.4  $ & $ 1.7$     & \multicolumn{2}{c|}{}       & 600  \\
                                     & 5282.285  & $-3190$ & $120$  (z)   & $-5.7  $ & $ 2.4$     & \multicolumn{2}{c|}{}       & 600  \\
                                     &           & $-3080$ & $50$   (r)   & \multicolumn{2}{c|}{} & \multicolumn{2}{c|}{}       &      \\
                                     & 5283.330  & $+3520$ & $170$  (z)   & $-8.6  $ & $ 3.7$     & \multicolumn{2}{c|}{}       & 500  \\
                                     &           & $+3350$ & $40$   (r)   & \multicolumn{2}{c|}{} & \multicolumn{2}{c|}{}       &      \\
                                     & 5284.329  & $+3460$ & $170$  (z)   & $-2.4  $ & $ 2.1$     & \multicolumn{2}{c|}{}       & 500  \\
                                     &           & $+3030$ & $50$   (r)   & \multicolumn{2}{c|}{} & \multicolumn{2}{c|}{}       &      \\
                                     & 5311.318  & $-6450$ & $150$  (z)   & $-7.7  $ & $ 2.9$     & \multicolumn{2}{c|}{}       & 500  \\
                                     &           & $-3630$ & $250$  (h)   & \multicolumn{2}{c|}{} & \multicolumn{2}{c|}{}       &      \\
                                     & 5315.335  & $+3990$ & $90$   (z)   & $-2.1  $ & $ 2.0$     & \multicolumn{2}{c|}{}       & 400  \\
                                     &           & $+2450$ & $250$  (h)   & \multicolumn{2}{c|}{} & \multicolumn{2}{c|}{}       &      \\
                                     & 5348.260  & $+4050$ & $120$  (z)   & $-3.2  $ & $ 1.3$     & \multicolumn{2}{c|}{}       & 400  \\
                                     &           & $+3700$ & $40$   (r)   & \multicolumn{2}{c|}{} & \multicolumn{2}{c|}{}       &      \\
                                     & 5552.638  & $-5580$ & $190$  (z)   & $-11.7 $ & $ 2.9$     & \multicolumn{2}{c|}{}       & 400  \\
\it{HD\,89069}                       & 5202.658  &  $-400$ & $50$   (z)   & $-2.5  $ & $ 1.1$     & \multicolumn{2}{c|}{$< 20$} & 400  \\
                                     & 5311.381  &  $-720$ & $30$   (z)   & $-12.7 $ & $ 2.8$     & \multicolumn{2}{c|}{}       & 300  \\
                                     &           &  $-260$ & $240$  (h)   & \multicolumn{2}{c|}{} & \multicolumn{2}{c|}{}       &      \\
                                     & 5315.395  &  $-250$ & $20$   (z)   & $-7.8  $ & $ 2.9$     & \multicolumn{2}{c|}{}       & 200  \\
                                     &           &  $-260$ & $30$   (r)   & \multicolumn{2}{c|}{} & \multicolumn{2}{c|}{}       &      \\
HD\,90763                            & 5552.668  &  $+180$ & $80$   (z)   & $-26.0 $ & $ 3.8$     & $50 $ & $ 5$                & 400  \\
                                     &           &   $-40$ & $60$   (r)   & \multicolumn{2}{c|}{} & \multicolumn{2}{c|}{}       &      \\
                                     & 5553.592  &  $-120$ & $110$  (z)   & $-25.5 $ & $ 3.4$     & \multicolumn{2}{c|}{}       & 500  \\
                                     &           &  $+100$ & $100$  (r)   & \multicolumn{2}{c|}{} & \multicolumn{2}{c|}{}       &      \\
                                     & 5554.625  &  $+170$ & $70$   (z)   & $-32.7 $ & $ 3.6$     & \multicolumn{2}{c|}{}       & 500  \\
                                     &           &  $-220$ & $130$  (r)   & \multicolumn{2}{c|}{} & \multicolumn{2}{c|}{}       &      \\
HD\,93294                            & 5202.560  &   $-40$ & $50$   (z)   & $+25.7 $ & $ 2.6$     & \multicolumn{2}{c|}{$< 20$} & 400  \\
                                     &           &   $-30$ & $310$  (h)   & \multicolumn{2}{c|}{} & \multicolumn{2}{c|}{}       &      \\
                                     & 5348.248  &   $-90$ & $20$   (z)   & $+21.3 $ & $ 2.1$     & \multicolumn{2}{c|}{}       & 300  \\
                                     &           &   $-80$ & $20$   (r)   & \multicolumn{2}{c|}{} & \multicolumn{2}{c|}{}       &      \\
\it{HD\,96003}                       & 5345.256  &  $-210$ & $20$   (z)   & $-13.7 $ & $ 2.4$     & \multicolumn{2}{c|}{$< 20$} & 600  \\
HD\,97633                            & 5552.675  &   $+40$ & $20$   (z)   & $-4.9  $ & $ 2.5$     & $25 $ & $ 5$                & 1500 \\
                                     &           &   $-30$ & $70$   (r)   & \multicolumn{2}{c|}{} & \multicolumn{2}{c|}{}       &      \\
                                     & 5553.673  &   $+10$ & $20$   (z)   & $-1.0  $ & $ 2.9$     & \multicolumn{2}{c|}{}       & 1000 \\
                                     &           &   $-20$ & $40$   (r)   & \multicolumn{2}{c|}{} & \multicolumn{2}{c|}{}       &      \\
                                     & 5554.640  &  $+120$ & $30$   (z)   & $-7.5  $ & $ 2.6$     & \multicolumn{2}{c|}{}       & 1500 \\
                                     &           &   $+40$ & $50$   (r)   & \multicolumn{2}{c|}{} & \multicolumn{2}{c|}{}       &      \\
HD\,108506                           & 5553.708  &  $+370$ & $510$  (z)   & $0.0   $ & $ 1.6$     & $150 $ & $ 20$              & 400  \\
                                     &           &  $-560$ & $330$  (r)   & \multicolumn{2}{c|}{} & \multicolumn{2}{c|}{}       &      \\
                                     & 5554.633  &  $+160$ & $250$  (z)   & $-16.2 $ & $ 3.4$     & \multicolumn{2}{c|}{}       & 500  \\
                                     &           &   $+50$ & $230$  (r)   & \multicolumn{2}{c|}{} & \multicolumn{2}{c|}{}       &      \\
                                     & 5555.566  &   $-60$ & $510$  (z)   & $-5.0  $ & $ 3.9$     & \multicolumn{2}{c|}{}       & 400  \\
                                     &           &  $-580$ & $250$  (r)   & \multicolumn{2}{c|}{} & \multicolumn{2}{c|}{}       &      \\
HD\,110066                           & 5345.244  &  $-220$ & $10$   (z)   & $-13.9 $ & $ 2.7$     & \multicolumn{2}{c|}{$< 20$} & 500  \\
HD\,112413**                         & 5202.581  &  $-880$ & $50$   (z)   & $+7.1  $ & $ 2.4$     & \multicolumn{2}{c|}{}       & 3000 \\
                                     &           &  $-680$ & $70$   (h)   & \multicolumn{2}{c|}{} & \multicolumn{2}{c|}{}       &      \\
                                     & 5281.468  &  $+860$ & $80$   (z)   & $+1.9  $ & $ 0.8$     & \multicolumn{2}{c|}{}       & 3000 \\
                                     & 5345.230  &  $-820$ & $59$   (z)   & $-1.7  $ & $ 1.3$     & \multicolumn{2}{c|}{}       & 2000 \\
\it{HD\,113894}                      & 5553.645  &  $+990$ & $40$   (z)   & $+8.2  $ & $ 2.4$     & $23 $ & $ 3$                & 300  \\
                                     &           &  $+900$ & $30$   (r)   & \multicolumn{2}{c|}{} & \multicolumn{2}{c|}{}       &      \\
                                     & 5554.650  &  $+840$ & $40$   (z)   & $+1.6  $ & $ 2.6$     & \multicolumn{2}{c|}{}       & 300  \\
                                     &           &  $+870$ & $30$   (r)   & \multicolumn{2}{c|}{} & \multicolumn{2}{c|}{}       &      \\
                                     & 5555.579  &  $+760$ & $30$   (z)   & $+4.6  $ & $ 2.1$     & \multicolumn{2}{c|}{}       & 300  \\
                                     &           &  $+690$ & $60$   (r)   & \multicolumn{2}{c|}{} & \multicolumn{2}{c|}{}       &      \\
HD\,114125                           & 5282.42   &   $+20$ & $80$   (z)   & $-56.1 $ & $ 3.1$     & $25 $ & $ 5$                & 400  \\
                                     &           &   $-50$ & $40$   (r)   & \multicolumn{2}{c|}{} & \multicolumn{2}{c|}{}       &      \\
                                     & 5345.279  &   $+20$ & $50$   (z)   & $-59.7 $ & $ 2.4$     & \multicolumn{2}{c|}{}       & 300  \\
\it{HD\,118054}                      & 5553.658  &  $-420$ & $190$  (z)   & $-10.4 $ & $ 2.5$     & $65 $ & $ 5$                & 300  \\
                                     &           &     $0$ & $140$  (r)   & \multicolumn{2}{c|}{} & \multicolumn{2}{c|}{}       &      \\
                                     & 5554.662  &  $-580$ & $130$  (z)   & $-17.8 $ & $ 2.6$     & \multicolumn{2}{c|}{}       &      \\
                                     &           &  $-890$ & $210$  (r)   & \multicolumn{2}{c|}{} & \multicolumn{2}{c|}{}       &      \\
HD\,135297                           & 5282.405  &   $+50$ & $50$   (z)   & $-44.5 $ & $ 2.9$     & $27 $ & $ 5$                & 500  \\
                                     &           &   $+70$ & $30$   (r)   & \multicolumn{2}{c|}{} & \multicolumn{2}{c|}{}       &      \\
                                     & 5345.324  &  $-170$ & $30$   (z)   & $-41.3 $ & $ 2.6$     & \multicolumn{2}{c|}{}       & 400  \\
HD\,135679                           & 5555.591  & $+1120$ & $40$   (z)   & $+0.8  $ & $ 2.1$     & \multicolumn{2}{c|}{$< 20$} & 410  \\
                                     &           &  $+960$ & $24$   (r)   & \multicolumn{2}{c|}{} & \multicolumn{2}{c|}{}       &      \\
HD\,137909**                         & 5202.590  &  $+430$ & $50$   (z)   & $+9.2  $ & $ 1.1$     & \multicolumn{2}{c|}{$< 20$} & 1800 \\
                                     &           &  $+270$ & $70$   (h)   & \multicolumn{2}{c|}{} & \multicolumn{2}{c|}{}       &      \\
\it{HD\,138633}                      & 5282.470  &  $+310$ & $30$   (z)   & $-14.6 $ & $ 2.8$     & \multicolumn{2}{c|}{$< 20$} & 300  \\
                                     &           &  $+350$ & $20$   (r)   & \multicolumn{2}{c|}{} & \multicolumn{2}{c|}{}       &      \\
                                     & 5287.489  &  $-290$ & $20$   (z)   & $-4.7  $ & $ 2.1$     & \multicolumn{2}{c|}{}       & 300  \\
                                     &           &  $-210$ & $370$  (h)   & \multicolumn{2}{c|}{} & \multicolumn{2}{c|}{}       &      \\
\it{HD\,138777}                      & 5348.300  & $+2150$ & $60$   (z)   & $-46.3 $ & $ 2.7$     & $25 $ & $ 5$                & 250  \\
                                     &           & $+2000$ & $40$   (r)   & \multicolumn{2}{c|}{} & \multicolumn{2}{c|}{}       &      \\
HD\,149046                           & 5345.335  &  $-105$ & $150$  (z)   & $-28.3 $ & $ 3.8$     & $25 $ & $ 5$                & 250  \\
HD\,152107**                         & 5287.450  & $+1180$ & $40$   (z)   & $+7.2  $ & $ 2.1$     & $30 $ & $ 5$                & 1000 \\
                                     &           &  $+290$ & $300$  (h)   & \multicolumn{2}{c|}{} & \multicolumn{2}{c|}{}       &      \\
HD\,157740                           & 5555.669  &  $+130$ & $70$   (z)   & $+6.5  $ & $ 3.1$     & $25 $ & $ 5$                & 300  \\
                                     &           &  $+130$ & $60$   (r)   & \multicolumn{2}{c|}{} & \multicolumn{2}{c|}{}       &      \\
HD\,158450                           & 5287.518  & $-4280$ & $110$  (z)   & $-12.7 $ & $ 2.4$     & $27 $ & $ 5$                & 300  \\
                                     &           & $-2320$ & $420$  (h)   & \multicolumn{2}{c|}{} & \multicolumn{2}{c|}{}       &      \\
                                     & 5315.420  & $-4430$ & $150$  (z)   & $-16.4 $ & $ 2.8$     & \multicolumn{2}{c|}{}       & 250  \\
                                     &           & $-3640$ & $40$   (r)   & \multicolumn{2}{c|}{} & \multicolumn{2}{c|}{}       &      \\
                                     & 5345.367  &  $-100$ & $65$   (z)   & $-19.6 $ & $ 2.7$     & \multicolumn{2}{c|}{}       & 300  \\
                                     & 5348.435  & $-4480$ & $130$  (z)   & $-18.2 $ & $ 2.4$     & \multicolumn{2}{c|}{}       & 250  \\
                                     &           & $-3900$ & $50$   (r)   & \multicolumn{2}{c|}{} & \multicolumn{2}{c|}{}       &      \\
HD\,158974*                          & 5287.458  &   $+20$ & $10$   (z)   & $-23.6 $ & $ 2.7$     & $25 $ & $ 5$                & 400  \\
                                     &           &   $+60$ & $460$  (h)   & \multicolumn{2}{c|}{} & \multicolumn{2}{c|}{}       &      \\
                                     & 5345.385  &   $-20$ & $10$   (z)   & $-34.3 $ & $ 3.4$     & \multicolumn{2}{c|}{}       & 600  \\
                                     & 5406.468  &   $-10$ & $10$   (z)   & $-30.3 $ & $ 2.9$     & \multicolumn{2}{c|}{}       & 400  \\
HD\,168856                           & 5287.581  &  $+360$ & $190$  (z)   & $-21.8 $ & $ 2.6$     & \multicolumn{2}{c|}{}       & 300  \\
                                     &           &   $-70$ & $280$  (h)   & \multicolumn{2}{c|}{} & \multicolumn{2}{c|}{}       &      \\
                                     & 5315.562  & $-1110$ & $320$  (z)   & \multicolumn{2}{c|}{} & \multicolumn{2}{c|}{}       & 400  \\
                                     &           &  $-250$ & $180$  (r)   & \multicolumn{2}{c|}{} & \multicolumn{2}{c|}{}       &      \\
HD\,169191*                          & 5460.252  &   $-20$ & $10$   (z)   & $-17.0 $ & $ 2.8$     & \multicolumn{2}{c|}{$< 20$} & 500  \\
                                     & 5461.284  &  $+142$ & $10$   (z)   & $-22.0 $ & $ 2.9$     & \multicolumn{2}{c|}{}       & 500  \\
                                     & 5553.102  &     $0$ & $10$   (z)   & $-14.0 $ & $ 2.8$     & \multicolumn{2}{c|}{}       & 500  \\
                                     &           &   $+10$ & $10$   (r)   & \multicolumn{2}{c|}{} & \multicolumn{2}{c|}{}       &      \\
HD\,178308                           & 5348.348  &   $+20$ & $120$  (z)   & $-24.8 $ & $ 2.9$     & $75 $ & $ 10$               & 400  \\
                                     &           &  $-140$ & $160$  (r)   & \multicolumn{2}{c|}{} & \multicolumn{2}{c|}{}       &      \\
                                     & 5554.125  &  $+240$ & $260$  (z)   & $-25.7 $ & $ 2.6$     & \multicolumn{2}{c|}{}       & 400  \\
                                     &           &  $+130$ & $130$  (r)   & \multicolumn{2}{c|}{} & \multicolumn{2}{c|}{}       &      \\
HD\,178892                           & 5460.235  & $+4580$ & $200$  (z)   & $-10.9 $ & $ 2.7$     & $30 $ & $ 5$                & 400  \\
                                     & 5461.268  & $+2790$ & $170$  (z)   & $-17.2 $ & $ 3.8$     & \multicolumn{2}{c|}{}       & 300  \\
HD\,182180                           & 5348.500  & $-4000$ & $1200$ (z)   & $+5.4  $ & $ 2.8$     & $200 $ & $ 30$              & 500  \\
                                     &           &  $-750$ & $310$  (r)   & \multicolumn{2}{c|}{} & \multicolumn{2}{c|}{}       &      \\
HD\,198920                           & 5406.527  &   $+10$ & $20$   (z)   & $+18.7 $ & $ 3.4$     & $30 $ & $ 5$                & 300  \\
                                     &           &   $+10$ & $10$   (r)   & \multicolumn{2}{c|}{} & \multicolumn{2}{c|}{}       &      \\
HD\,199180                           & 5287.592  &  $-340$ & $60$   (z)   & $-17.3 $ & $ 2.8$     & $27 $ & $ 5$                & 300  \\
                                     &           &  $-350$ & $300$  (h)   & \multicolumn{2}{c|}{} & \multicolumn{2}{c|}{}       &      \\
HD\,201174                           & 5345.407  & $+1810$ & $90$   (z)   & $-13.8 $ & $ 3.4$     & $25 $ & $ 5$                & 400  \\
                                     & 5459.442  &  $+710$ & $90$   (z)   & $-10.6 $ & $ 2.7$     & \multicolumn{2}{c|}{}       & 300  \\
                                     & 5460.193  & $+2110$ & $50$   (z)   & $-4.6  $ & $ 2.8$     & \multicolumn{2}{c|}{}       & 300  \\
                                     & 5554.190  &  $+680$ & $100$  (z)   & $-16.5 $ & $ 3.4$     & \multicolumn{2}{c|}{}       & 300  \\
                                     &           &  $+480$ & $60$   (r)   & \multicolumn{2}{c|}{} & \multicolumn{2}{c|}{}       &      \\
                                     & 5555.159  & $+2070$ & $80$   (z)   & $-9.9  $ & $ 3.4$     & \multicolumn{2}{c|}{}       & 350  \\
                                     &           & $+1930$ & $40$   (r)   & \multicolumn{2}{c|}{} & \multicolumn{2}{c|}{}       &      \\
HD\,201601**                         & 5287.600  & $-1190$ & $30$   (z)   & $-18.3 $ & $ 2.0$     & \multicolumn{2}{c|}{$< 20$} & 1000 \\
                                     &           &  $-630$ & $460$  (h)   & \multicolumn{2}{c|}{} & \multicolumn{2}{c|}{}       &      \\
                                     & 5431.480  & $-1230$ & $30$   (z)   & $-14.9 $ & $ 2.7$     & \multicolumn{2}{c|}{}       & 700  \\
                                     & 5460.218  & $-1050$ & $30$   (z)   & $-20.3 $ & $ 2.9$     & \multicolumn{2}{c|}{}       & 1000 \\
                                     & 5461.250  & $-1070$ & $50$   (z)   & $-19.1 $ & $ 2.8$     & \multicolumn{2}{c|}{}       & 1000 \\
                                     & 5488.357  & $-1020$ & $50$   (z)   & $-14.4 $ & $ 3.8$     & \multicolumn{2}{c|}{}       & 600  \\
                                     & 5553.160  & $-1130$ & $30$   (z)   & $-16.6 $ & $ 2.7$     & \multicolumn{2}{c|}{}       & 900  \\
                                     &           & $-1000$ & $30$   (r)   & \multicolumn{2}{c|}{} & \multicolumn{2}{c|}{}       &      \\
                                     & 5554.100  & $-1050$ & $40$   (z)   & $-16.3 $ & $ 2.6$     & \multicolumn{2}{c|}{}       & 1000 \\
                                     & 5555.100  & $-1120$ & $40$   (z)   & $-16.7 $ & $ 2.4$     & \multicolumn{2}{c|}{}       & 1000 \\
                                     &           &  $-960$ & $30$   (r)   & \multicolumn{2}{c|}{} & \multicolumn{2}{c|}{}       &      \\
HD\,217401                           & 5488.381  &   $-30$ & $70$   (z)   & $+1.4  $ & $ 2.6$     & $55 $ & $ 10$               & 350  \\
                                     &           &   $-80$ & $70$   (r)   & \multicolumn{2}{c|}{} & \multicolumn{2}{c|}{}       &      \\
                                     & 5554.240  &  $-140$ & $90$   (z)   & $-9.0  $ & $ 3.1$     & \multicolumn{2}{c|}{}       & 250  \\
                                     &           &   $-30$ & $50$   (r)   & \multicolumn{2}{c|}{} & \multicolumn{2}{c|}{}       &      \\
HD\,225627                           & 5348.470  &  $+340$ & $60$   (z)   & $+16.3 $ & $ 2.8$     & $30 $ & $ 5$                & 300  \\
                                     &           &  $+320$ & $50$   (r)   & \multicolumn{2}{c|}{} & \multicolumn{2}{c|}{}       &      \\
                                     & 5554.162  &   $-60$ & $40$   (z)   & $+11.1 $ & $ 3.1$     & \multicolumn{2}{c|}{}       & 300  \\
                                     &           &   $-20$ & $40$   (r)   & \multicolumn{2}{c|}{} & \multicolumn{2}{c|}{}       &      \\
HD\,258686                           & 5202.304  & $+5750$ & $320$  (z)   & $+15.4 $ & $ 2.1$     & $35 $ & $ 5$                & 300  \\
                                     &           & $+6880$ & $450$  (h)   & \multicolumn{2}{c|}{} & \multicolumn{2}{c|}{}       &      \\
HD\,261937                           & 5554.556  &  $-720$ & $970$  (z)   & $+16.0 $ & $ 3.1$     & $130 $ & $ 10$              & 200  \\
                                     &           &  $+420$ & $100$  (r)   & \multicolumn{2}{c|}{} & \multicolumn{2}{c|}{}       &      \\
                                     & 5555.523  & $-1050$ & $1460$ (z)   & $+15.8 $ & $ 3.4$     & \multicolumn{2}{c|}{}       & 250  \\
                                     &           &  $-150$ & $70$   (r)   & \multicolumn{2}{c|}{} & \multicolumn{2}{c|}{}       &      \\
HD\,279021                           & 5554.458  & $+1040$ & $90$   (z)   & $+1.3  $ & $ 1.2$     & $35 $ & $ 5$                & 250  \\
                                     &           &  $+860$ & $120$  (r)   & \multicolumn{2}{c|}{} & \multicolumn{2}{c|}{}       &      \\
                                     & 5555.375  &  $+470$ & $90$   (z)   & $+2.9  $ & $ 1.2$     & \multicolumn{2}{c|}{}       & 250  \\
                                     &           &  $+490$ & $70$   (r)   & \multicolumn{2}{c|}{} & \multicolumn{2}{c|}{}       &      \\
HD\,281367                           & 5555.416  & $+1250$ & $1100$ (z)   & $+46.9 $ & $ 3.7$     & $45 $ & $ 5$                & 200  \\
                                     &           &  $-140$ & $100$  (r)   & \multicolumn{2}{c|}{} & \multicolumn{2}{c|}{}       &      \\
HD\,343872                           & 5281.527  & $+2870$ & $140$  (z)   & $-2.9  $ & $ 1.4$     & $25 $ & $ 5$                & 200  \\
                                     & 5282.508  & $+1900$ & $160$  (z)   & \multicolumn{2}{c|}{} & \multicolumn{2}{c|}{}       & 250  \\
                                     &           & $+1380$ & $30$   (r)   & \multicolumn{2}{c|}{} & \multicolumn{2}{c|}{}       &      \\
                                     & 5283.524  &     $0$ & $100$  (z)   & $-10.3 $ & $ 2.4$     & \multicolumn{2}{c|}{}       & 300  \\
                                     & 5284.604  &  $-820$ & $150$  (z)   & $-5.6  $ & $ 2.4$     & \multicolumn{2}{c|}{}       & 160  \\
                                     &           &  $-930$ & $30$   (r)   & \multicolumn{2}{c|}{} & \multicolumn{2}{c|}{}       &      \\
                                     & 5287.552  & $+4340$ & $90$   (z)   & $-3.7  $ & $ 2.0$     & \multicolumn{2}{c|}{}       & 200  \\
                                     &           & $+4420$ & $410$  (h)   & \multicolumn{2}{c|}{} & \multicolumn{2}{c|}{}       &      \\
                                     & 5348.395  & $+4280$ & $120$  (z)   & $-4.0  $ & $ 2.8$     & \multicolumn{2}{c|}{}       & 200  \\
                                     &           & $+3760$ & $40$   (r)   & \multicolumn{2}{c|}{} & \multicolumn{2}{c|}{}       &      \\
{\it $BD\,+{\it 53}\degr{\it 1183}$} & 5202.630  &  $-110$ & $120$  (z)   & $+8.4  $ & $ 1.4$     & $50 $ & $ 5$                & 250  \\
                                     & 5311.347  & $+1030$ & $100$  (z)   & $+0.3  $ & $ 2.1$     & \multicolumn{2}{c|}{}       & 250  \\
                                     &           &  $-280$ & $510$  (h)   & \multicolumn{2}{c|}{} & \multicolumn{2}{c|}{}       &      \\
                                     & 5315.365  &  $-810$ & $110$  (z)   & $+0.9  $ & $ 2.8$     & \multicolumn{2}{c|}{}       & 250  \\
                                     &           &  $-510$ & $79$   (r)   & \multicolumn{2}{c|}{} & \multicolumn{2}{c|}{}       &      \\
${\rm BD}\,+38\degr2360$             & 5555.635  &   $-50$ & $70$   (z)   & $-14.8 $ & $ 2.9$     & $45 $ & $ 5$                & 300  \\
                                     &           &   $+70$ & $50$   (r)   & \multicolumn{2}{c|}{} & \multicolumn{2}{c|}{}       &      \\
${\rm BD}\,+37\degr431$              & 5555.169  &   $+10$ & $30$   (z)   & $-10.3 $ & $ 1.6$     & $30 $ & $ 5$                & 250  \\
                                     &           &     $0$ & $40$   (r)   & \multicolumn{2}{c|}{} & \multicolumn{2}{c|}{}       &      \\
${\rm BD}\,+36\degr363$              & 5488.538  &  $+380$ & $480$  (z)   & $+9.8  $ & $ 2.9$     & $100 $ & $ 15$              & 250  \\
                                     &           &   $+50$ & $100$  (r)   & \multicolumn{2}{c|}{} & \multicolumn{2}{c|}{}       &      \\
${\rm BD}\,+00\degr4535$             & 5406.499  &  $-570$ & $130$  (z)   & $-49.8 $ & $ 3.4$     & $23 $ & $ 5$                & 200  \\
                                     &           &  $-250$ & $40$   (r)   & \multicolumn{2}{c|}{} & \multicolumn{2}{c|}{}       &      \\
                                     & 5553.130  & $-1990$ & $120$  (z)   & $-36.1 $ & $ 3.6$     & \multicolumn{2}{c|}{}       & 200  \\
                                     &           & $-1410$ & $60$   (r)   & \multicolumn{2}{c|}{} & \multicolumn{2}{c|}{}       &      \\
${\rm BD}\,-12\degr2366$             & 5552.522  &  $-100$ & $100$  (z)   & $+32.9 $ & $ 2.8$     & $40 $ & $ 5$                & 170  \\
                                     &           &   $+20$ & $40$   (r)   & \multicolumn{2}{c|}{} & \multicolumn{2}{c|}{}       &      \\
                                     & 5553.635  &   $+70$ & $70$   (z)   & $+37.9 $ & $ 2.7$     & \multicolumn{2}{c|}{}       & 170  \\
                                     &           &  $+100$ & $80$   (r)   & \multicolumn{2}{c|}{} & \multicolumn{2}{c|}{}       &      \\
NGC\,752-105                         & 5555.215  &  $-970$ & $620$  (z)   & $-5.9  $ & $ 3.1$     & $75 $ & $ 10$               & 150  \\
$o$\,UMa*                            & 5281.460  &   $-30$ & $20$   (z)   & $+23.7 $ & $ 2.1$     & $23 $ & $ 5$                & 400  \\
                                     & 5282.294  &   $+90$ & $20$   (z)   & $+18.6 $ & $ 3.4$     & \multicolumn{2}{c|}{}       & 500  \\
                                     &           &  $+110$ & $10$   (r)   & \multicolumn{2}{c|}{} & \multicolumn{2}{c|}{}       &      \\
                                     & 5283.340  &  $-110$ & $30$   (z)   & $+14.4 $ & $ 2.4$     & \multicolumn{2}{c|}{}       & 800  \\
                                     &           &  $-140$ & $10$   (r)   & \multicolumn{2}{c|}{} & \multicolumn{2}{c|}{}       &      \\
                                     & 5284.336  &     $0$ & $10$   (z)   & $+20.7 $ & $ 3.4$     & \multicolumn{2}{c|}{}       & 800  \\
                                     &           &     $0$ & $10$   (r)   & \multicolumn{2}{c|}{} & \multicolumn{2}{c|}{}       &      \\
                                     & 5311.311  &   $-20$ & $10$   (z)   & $+22.0 $ & $ 2.8$     & \multicolumn{2}{c|}{}       & 1300 \\
                                     & 5315.328  &   $-40$ & $10$   (z)   & $+21.2 $ & $ 3.7$     & \multicolumn{2}{c|}{}       & 1300 \\
                                     & 5348.268  &     $0$ & $10$   (z)   & $+19.7 $ & $ 2.7$     & \multicolumn{2}{c|}{}       & 900  \\
                                     &           &     $0$ & $10$   (r)   & \multicolumn{2}{c|}{} & \multicolumn{2}{c|}{}       &      \\
                                     & 5459.470  &   $-50$ & $50$   (z)   & $+17.0 $ & $ 2.8$     & \multicolumn{2}{c|}{}       & 800  \\
                                     & 5552.644  &  $-190$ & $10$   (z)   & $+15.8 $ & $ 1.8$     & \multicolumn{2}{c|}{}       & 1000 \\
                                     &           &  $-240$ & $20$   (r)   & \multicolumn{2}{c|}{} & \multicolumn{2}{c|}{}       &      \\
 \hline

\end{longtable*}
\twocolumngrid

\section {COMMENTS}

This Section gives comments and notes to the data on individual
stars. If they were observed by us earlier and have already been
described in papers~\mbox{\cite{1:Romanyuk_n_en_4,2:Romanyuk_n_en_4,3:Romanyuk_n_en_4}}, the corresponding reference
is given. More attention is given to the stars which were observed
for the first time  in 2010. We preserve the traditional sequence of
comments adopted in the previous articles.

\subsection {Non-Magnetic  Standard Stars}

 We use cool slowly
rotating stars with a lot of narrow lines as non-magnetic  standard stars.
There is no strong general magnetic field in such objects
and the accuracy of field measurements is very high.

\subsubsection {HD\,33256}

The star of spectral class F5. We use it as a
zero standard. As it can be seen from the Table,  the scattering
did not exceed 50~G in any of the cases, which fully
corresponds to the expected accuracy.

\subsubsection {HD\,71369 = $o$\,UMa}

A zero standard. Ten field measurements were obtained. With the
exception of one night \linebreak (JD=2455552.644), systematic errors
are within
 100~G. It is possible that a large deviation from zero in
the given night is related to a poor  adjustment of the new
 large CCD in the initial period of its operation.

\subsubsection {HD\,158974}

A zero standard. In 2010, systematic measurement errors did not
exceed 16~G. The details are presented in the papers dedicated to the
results of
\mbox{2007--2009}~\cite{1:Romanyuk_n_en_4,2:Romanyuk_n_en_4,3:Romanyuk_n_en_4}.

\subsubsection  {HD\,169191}

A zero standard. We see that in 2010 the systematic
measurement errors did not exceed 20~G, except for one case (JD =
2455461.284) when they amounted to $142\pm10$~G. Apparently, there have
been errors
of star guiding captured in the slit of the spectrograph. It could leave the
slit at short exposures, and the observers did not have time to
compensate for this departure.

Measurements of zero standards show that in general there are no
systematic errors that could lead to a distortion of the
results. Nevertheless, in some cases there are more
significant deviations. Hence, if observations show that
the star has a  $B_e$ field smaller than 100~G, we do not consider
the star to be magnetic
even at very small formal measurement errors.

\subsection {Magnetic Standard  Stars}

Chemically peculiar
stars with reliably determined longitudinal field  $B_e$ variability curves
 are selected as the magnetic standards.

\subsubsection {HD\,65339 = 53\,Cam}

One of the most well-studied magnetic CP stars. We observe it
systematically for the purpose of standardization
of observations~\cite{1:Romanyuk_n_en_4,2:Romanyuk_n_en_4,3:Romanyuk_n_en_4}. In 2010, nine
observations of this object were performed. Our results can be used
when studying the long-term variability of the star.

\subsubsection {HD\,112413 = $\alpha^2$\,Cvn}

The star with a well-known curve of the longitudinal field, the brightest
of all known magnetic CP stars. It is observed for
calibration. As before, the results of our measurements
of the field correspond to the ephemeris.
Dozens of works, devoted to the study of this star are annually published,
including
detailed investigations of its magnetic field.

\subsubsection {HD\,137909= $\beta$\,CrB}

$\beta$\,CrB is a very well studied, second in brightness
magnetic CP star. It is used to calibrate the data
as the standard of the magnetic field.

\subsubsection {HD\,152107 = 52\,Her}

52\,Her is a  well-studied CP star, convenient for the purposes
 of calibration, since the longitudinal component of the field $B_e$ has
a constant positive polarity. In the 2010 observations a
good agreement was obtained with the data of previous studies.
The star is a binary, with the orbital period of about 56~years~\cite{5:Romanyuk_n_en_4},
therefore, we observe  variability of its radial velocity.

\subsubsection {HD\,201601 = $\gamma$\,Equ}

$\gamma$ Equ is a magnetic star with the longest rotation period
 (about 100~years). In 2010, eight measurements of the
field were performed, the average value of $B_e$ over the year is
$-1107\pm26$~G (by the classical Babcock method). This means that due
to rotation its longitudinal field $B_e$ begins to depart from the
phase of the negative extremum. The estimate of the field by the
regression method is somewhat smaller than the value obtained in the
classical way. Systematic radial velocity variations in comparison
with 2009 are not noted. The average value of  \mbox
{$V_r=-17.1\pm0.7$}~km\,s$^{-1}$  within the errors coincides with
the SIMBAD estimate of  $V_r=-16.5$~km\,s$^{-1}$.

\subsection {Stars with Magnetic Field   Registered in 2010}

\subsubsection {HD\,965}

A very slowly rotating magnetic star of the SrCrEu-type. The period of its
rotation reaches 20~years. Magnetic  monitoring of the star is conducted
by us regularly for more than 15~years. Preliminary
 results are published in~\cite{6:Romanyuk_n_en_4}. In 2010, the longitudinal
field $B_e$  passed from the region of positive values to the region of
negative values. The average  $B_e$ over a year is
\mbox {$-230$}~G.  Radial velocity of the star is approximately the same as
  in 2009. Note that the paper for 2009 contains an annoying
typo: on the date JD=2455075.438   radial velocity is
$-5.8$~km\,s$^{-1}$, instead of \mbox {$+9.8$}~km\,s$^{-1}$,
as
erroneously printed in the specified article. No signs of HD\,965 binarity
   were  ever found.

\subsubsection {HD\,5441}

SrCrEu is a star of spectral class A2. The first measurements of its
magnetic field were performed by us in 2009. One measurement in 2010
gave the same value of $B_e= -440\pm30$~G. The lines are very narrow
and sharp. The classical Zeeman and the regression method give the
same field value. For more details about the star, see~\cite{3:Romanyuk_n_en_4}. The
GAIA parallax $\pi = 2.60$~mas, the variable
  radial velocity of the star is +43~km\,s$^{-1}$ (in comparison with
+18~km\,s$^{-1}$ and +37~km\,s$^{-1}$  in 2009) points  to binarity.
Therefore, a new magnetic binary star is discovered.

\subsubsection {HD\,5797}

The magnetic field of this star was discovered by us earlier (see
paper~\cite{7:Romanyuk_n_en_4}). Semenko et al. investigated it in the work~\cite{8:Romanyuk_n_en_4}.
The radial velocity, measured in 2010, was about $-3.5$~km\,s$^{-1}$,
it systematically differs from that measured in 2009. We confirm that
the star is binary.

\subsubsection {HD\,6757}

A magnetic star with a strong  depression in the continuum. The history of its
study is described in detail in the papers~\cite{1:Romanyuk_n_en_4,2:Romanyuk_n_en_4}. In~\cite{7:Romanyuk_n_en_4}, we reported the discovery of a field. One
the discovery of a field in it. One 2010 measurement
  (+2800~G) confirms the presence of a strong field,
whose longitudinal component has a constant positive
polarity with a weak variability relative to the mean value
(about +2700~G). The star is a multiple system, the main component of
which has a magnetic field. It is studied in more detail in the recent
work~\cite{9:Romanyuk_n_en_4}.

\subsubsection  {HD\,16705}

A chemically peculiar star with weakened helium lines~\cite{10:Romanyuk_n_en_4}
in the scattered  NGC\,1039 cluster aged $\log t =8.26$  (according to
the VIZIER database).

An attempt to measure the magnetic field was performed for the first
time. Owing to  very wide lines with complex profiles, the  $B_e$
value could not be reliably measured. The presence of the field can
only be suspected. We estimate the projection of the rotation
velocity onto the line of sight as \mbox{$v_e\sin i
=100$}~km\,s$^{-1}$. The  VIZIER database  in several   quoted
publications lists the rotation period of the star as \mbox {$P =
9\fd 944$}, but we consider it to be erroneous. Since the equatorial
  rotation velocity of the star is at least 100~km\,s$^{-1}$, its period of
rotation can not exceed two days. In 2010, radial velocity amounted
to \mbox {$V_r= -$12.2}~km\,s$^{-1}$. Multiple catalogs show the
presence of a faint companion   at a distance of about   $20\arcsec$.
Apparently, this is an visual pair.

\subsubsection {HD\,17330}

A new magnetic star. The first attempt to measure the field was
carried out in 2010. The star was suspected of being magnetic, since
the field was found only by the regression method, and the classical
method did not allow  to discover it. The measurements made in
subsequent years confirmed the presence of a magnetic field in the
star, the longitudinal component of which $B_e$ reaches $-400$~G.
According to the Renson and Manfroid
catalog~\cite{10:Romanyuk_n_en_4},  the silicon lines are
strengthened in the star of class B7. The SIMBAD database shows the
radial velocity of the star of \mbox{$V_r=-2.5$~km\,s$^{-1}$};  it
differs from the one  we measured,
 $-13.6$~km\,s$^{-1}$. It is noted that
the star is binary, and a companion of 11th magnitude is at a distance of
$10\arcsec$. The lines in the spectrum are very narrow, the value of
 $v_e \sin i$ does not
exceed 20~km\,s$^{-1}$. Most likely, the observer sees the star
at a small angle $i$.

\subsubsection  {HD\,29762}

A new magnetic star. A large number of narrow
and sharp lines is observed in the spectrum, and hence
high-accuracy measurements can be performed.

\subsubsection {HD\,35298}

Borra~\cite{11:Romanyuk_n_en_4} discovered this star to be magnetic. In 2010 our
first Zeeman spectra of the star were obtained. Further, the
observations were continued, the longitudinal field variability curve
and the magnetic star model were constructed~\cite{12:Romanyuk_n_en_4}. The results
of measurements strongly depend on the technique used. Regression
analysis gives the field two times smaller than the one yielded by
the classical Babcock method. The SIMBAD database lists the radial
velocity of $V_r = +30$~km\,s$^{-1}$,  which is close to the one we
found. There is no data about the star's binarity.

\subsubsection {HD\,35379}

A chemically peculiar star. The catalog~\cite{10:Romanyuk_n_en_4} shows
a SiSr peculiarity. From one 2010 measurement
the presence of a magnetic field can be suspected.
The star is poorly studied, other observations
were not performed. There are many lines in the spectrum:
 \mbox{$v_e \sin i =
45$}~km\,s$^{-1}$, $V_r = +3.7$~km\,s$^{-1}$.

\subsubsection  {HD\,35456}

A well-known magnetic star. Borra found its field~\cite{11:Romanyuk_n_en_4}.
In 2010 the field was also detected by us. Subsequently, we
continued the observations of HD\,35456; the results of our field measurements
are published in the paper by Romanyuk et al.~\cite{13:Romanyuk_n_en_4}.

\subsubsection  {HD\,35881}

The star with weakened lines of helium, a member of the Orion\,OB1a association.
Rotation is very fast: $v_e \sin i=200$~km\,s$^{-1}$.
The first observations  with a Zeeman analyzer were made in 2010.
The accuracy of the field determination is very low. Subsequently, observations
were continued and the results were published in Romanyuk et al.~\cite{13:Romanyuk_n_en_4}.
We suspected that the star is magnetic.

\subsubsection {HD\,36313}

A previously known magnetic star. The field was discovered by Borra~\cite{11:Romanyuk_n_en_4}
using the Balmer magnetometer. However, our 2010 observation
  performed by the narrow metal lines did not reveal the presence of a field
   (see the Table). Further observations confirmed that
the star is magnetic. The main component, a fast
rotator possesses the field:  its spectrum reveals several lines strongly
broadened
by rotation. Narrow lines in the spectrum belong to the secondary
component, to a~cooler non-magnetic star, a slow
rotator. Details are described in Romanyuk et al.~\cite{13:Romanyuk_n_en_4}. A member of the
Orion\,OB1 association. Using speckle interferometry, Balega et al.~\cite{14:Romanyuk_n_en_4}
discovered a companion  at a distance of  $0\farcs15$.

\subsubsection {HD\,36485}

We conducted six field measurements (two in each of the three in a row
December nights in 2010). The field of negative polarity is  20\%
weaker when measured by the regression method in comparison with
the Babcock classical method. The star with strengthened helium lines,
the member of the Orion\,OB1b association, a spectral binary.
The SIMBAD database gives
the radial velocity of   $V_r=+21$~km\,s$^{-1}$,   which is close to
our results.
The field we found is somewhat smaller than
the estimate given in the literature when measured by Landstreet's
hydrogen magnetometer~\cite{15:Romanyuk_n_en_4}.

\subsubsection {HD\,36526}

Our 2010 measurement confirmed the presence of a very strong
magnetic field of the star, described in the paper~\cite{11:Romanyuk_n_en_4}.
Subsequently, we performed another set of measurements and determined the period
of  rotation of HD\,36526. The results are published in  Romanyuk et al.~\cite{13:Romanyuk_n_en_4}.

\subsubsection {HD\,36540}

The star with weakened lines of helium. The magnetic field is detected,
which confirmed the result of Borra~\cite{11:Romanyuk_n_en_4}. The measurements
were further continued. There is information indicating that the field
of the star
is rather weak, the longitudinal component does not exceed 1~kG. The results of a
detailed study of the star were published in Romanyuk et al.~\cite{16:Romanyuk_n_en_4}.

\subsubsection  {HD\,36916}

The magnetic field of the star was discovered by Borra et al.~\cite{17:Romanyuk_n_en_4} during
a large magnetic survey of stars with weakened
helium lines. We confirmed the presence of the field. Our subsequent measurements
allowed to investigate the variability of the longitudinal component of the
field~\cite{16:Romanyuk_n_en_4}.

\subsubsection {HD\,37022}

A hot star (spectral class O7) is a member of the multiple system
$\theta^1$\,Ori\,C. We performed six field measurements. The value of
the measured field, up to 500~G, is in agreement with the results
published earlier in the literature~\cite{18:Romanyuk_n_en_4}. Lines in the spectrum
have a very complex profile. Field measurements are difficult. Radial velocities
are variable. The star is magnetic, although not formally included in
the list of peculiar stars by Renson and Manfroid~\cite{10:Romanyuk_n_en_4}. A
moderate linear polarization (about 0.4\%) is observed, bound,
apparently, with the fact that the star is in the Great
Orion Nebula.

\subsubsection {HD\,37140}

The longitudinal magnetic field of the star was detected by Borra~\cite{11:Romanyuk_n_en_4}. It
is  variable~in the range from   $-1050$ to $+400$~G.
Our single measurement of 2010 gives a value of $B_e$ of about
  $-500$~G
by the Babcock method and $-400$~G by the regression method. Radial
  velocity in the SIMBAD database is  \mbox{$V_r = +14.7$~km\,s$^{-1}$},
  which   significantly
differs from the value we obtained.
In the Washington Observatory catalog of binary stars
 HD\,37140 is represented as a binary. Its companion is found at the
distance of $0\farcs 1$.
The lines in the spectrum are narrow and
sharp, which allows performing high-precision measurements of the magnetic
field of the star.

\subsubsection {HD\,37479}

The known magnetic star $\sigma$\,Ori\,E. In the paper~\cite{19:Romanyuk_n_en_4} 22
measurements were made on a hydrogen magnetometer based on the
H$_{\beta}$ and He~$\lambda\,5876$ lines. A curve was constructed
with the: \\ ${\rm JD (min)} = 2442778.819 + 1\fd19801$,\\ \mbox
{$B_{1} = 2150 \pm 120$}~G,\\ \mbox {$B_{0}= 660 \pm 60$}~G,\\
$\phi_0= 0.474$ elements.
  One of our 2010 measurements did not show the presence of a
field, wide lines did not allow it to be measured based on metals.
The measurements from the hydrogen lines are required.

\subsubsection {HD\,37687}

The star with weakened lines of helium, a member of the Orion\,OB1c
association. One observation yielded a longitudinal field of about
+500~G in full agreement with the results of~\cite{20:Romanyuk_n_en_4} obtained in
2004 and 2005.

\subsubsection {HD\,37776}

In 2010, two observations of this unique magnetic
star were performed. The field measured by the Babcock method
dramatically differs from that obtained by the regression method.
The star was observed with the
calibration aim. It is possible to see more  details on this unique object
in Kochukhov et al.~\cite{21:Romanyuk_n_en_4}. Radial velocity within the
errors  coincides with that  given in the SIMBAD database, what
testifies in favor of the fact that the star is single.

\subsubsection {HD\,38823}

The magnetic field was detected by us at the observations on the
BTA~\cite{7:Romanyuk_n_en_4}. The longitudinal component varies from $-2500$ to
$+1500$~G, however, the curve has not yet been   constructed. The
observation of 2010 was at the phase when $B_e$ was close to zero.
The radial velocity within the error   coincides with that indicated
in the SIMBAD database.

\subsubsection {HD\,45583}

We discovered the magnetic field of the star (see~\cite{7:Romanyuk_n_en_4}), in
detail the history of its magnetic research is given in the
paper~\cite{3:Romanyuk_n_en_4}. The longitudinal component varies with a large
amplitude, the variability curve is not sinusoidal. The radial
velocity reveals a weak variability. The SIMBAD database lists the
value \mbox{$V_r = +22.8$~km\,s$^{-1}$}.

\subsubsection {HD\,49884}

A new magnetic star. The spectral class A0, the Sr-type peculiarity.
Three 2010 measurements showed the presence of a weak field, the
longitudinal
 component of which has a negative polarity. Lines are very
narrow. The radial velocity might be variable.

\subsubsection {HD\,50169}

One measurement of 2010 showed that the longitudinal field of this
super-slow  rotator was in the phase of the period when
the magnetic equator of the star is observed. The transition from
the negative to   positive polarity of the longitudinal field is visible.
A more detailed  history of studying the magnetic field of the object is
described in~\cite{1:Romanyuk_n_en_4}. The radial velocity \mbox {$V_r =
+14.6$}~km\,s$^{-1}$ is in complete agreement with the data
 presented in the
SIMBAD database \mbox{($V_r = +13.2$~km\,s$^{-1}$)}.

\subsubsection {HD\,50461}

The star has a large value of the photometric index  $\Delta a =
0.052$. We detected a magnetic field (see~\cite{7:Romanyuk_n_en_4}), the results
of the subsequent measurements are presented in~\cite{2:Romanyuk_n_en_4}. One 2010 measurement
 showed the presence of a field of positive polarity. The lines in
the spectrum are broadened by rotation, hence it was impossible to achieve a
 high accuracy of measurements. The SIMBAD database
gives \mbox {$V_r = +38.1$}~km\,s$^{-1}$,  which differs from the
value we found \mbox{$V_r = +29.5$~km\,s$^{-1}$}.

\subsubsection {HD\,51418}

A rare magnetic star with holmium  and dysprosium
anomalies~\cite{22:Romanyuk_n_en_4}. Four measurements, carried out
in 2010, revealed the presence of a magnetic field. The elements of
photometric variability are  \mbox{${\rm JD} ({\rm max}~V) =
2441241.654 + 5\fd4379$}. The magnetic maximum coincides with the
maximum brightness. The results of our    radial velocity
measurements systematically differ from those presented in the SIMBAD
database: \mbox{$ V_r = -22.5 $}~km\,s $ ^ {- 1} $. In the work of
Balega et al.~\cite{14:Romanyuk_n_en_4}   the star revealed a
companion at a distance of $0\farcs 15$, weaker by three magnitudes.

\subsubsection {HD\,54824}

A new magnetic star discovered by us in 2010. All three
measurements showed the presence of a magnetic field. A binary star
ADS\,5852AB with strengthened strontium lines. Radial velocity is
variable, $v_e \sin i = 50$~km\,s$^{-1}$.

\subsubsection {HD\,89069}

A new magnetic star, the field is visible on all three spectra of
2010. According to the catalog of Renson and
Manfroid~\cite{10:Romanyuk_n_en_4}, it has SrCrEu peculiarities. The
lines in the spectrum are very narrow, the accuracy of field
measurements is high. The period of rotation of the star is about
18~days~\cite{23:Romanyuk_n_en_4}. Speckle-interferometry allowed to
resolve a companion~\cite{24:Romanyuk_n_en_4} at a distance of
$3\farcs 5$.  Our observations showed that  radial velocities are
variable, but the estimates differ from \mbox{$V_r =
-10.7$~km\,s$^{-1}$} (SIMBAD).

\subsubsection {HD\,96003}

A new magnetic star. One measurement of 2010 showed the presence of
a weak field of negative polarity. The information about the magnetic
 field of the star was not previously published. Further studies confirmed
the presence of a magnetic field~\cite{9:Romanyuk_n_en_4} \mbox {$V_r =
-1.1$}~km\,s$^{-1}$ (SIMBAD), which is quite different from ours
values. The star is binary and magnetic.

\subsubsection {HD\,110066}

A monitoring of the magnetic field of this very long-period
star was performed. One 2010 measurement gave the value of the longitudinal
magnetic field of about \mbox{$-200$}~G. The results of previous
measurements are given in Romanyuk et al.~\cite{1:Romanyuk_n_en_4}.

\subsubsection {HD\,113894}

We have discovered a new magnetic star. All the three 2010 measurements
revealed a field of positive polarity. The lines in the spectrum are very
narrow, the accuracy of measurements is high.
Renson and
Manfroid~\cite{10:Romanyuk_n_en_4} mark a SrCrEu peculiarity. The rotation period
of the star is at least 10~days. The observations of the star were continued. The
SIMBAD database lists the value of $V_r = +10$~km\,s$^{-1}$, which
is quite different from what we measured. Our
measurements, carried out for three consecutive nights reveal a
significant scatter in radial velocity determination, but
nonetheless do not go beyond the limits of formal errors. The mean out of three
 values we obtained is  \mbox {$+4.8 \pm 1.9$}~km\,s$^{-1}$.
There is a suspicion that the system is binary.

\subsubsection {HD\,118054}

A new magnetic star discovered by us in 2010. The field $B_e$
is smaller than 1~kG, of negative polarity. We determined the value
of the projection of the rotation velocity onto the line of sight
 \mbox{$v_e \sin i= 65$}~km\,s$^{-1}$. According to the catalog~\cite{10:Romanyuk_n_en_4}, the star is of
the SrEuCr peculiarity type. A visual binary of ADS\,8954. The SIMBAD database lists
$V_r=-14.4$~km\,s$^{-1}$, which corresponds to our measurements.

\subsubsection {HD\,138633}

A new magnetic star, the field was discovered by us for the first
time in 2010. The SrCrEu-type peculiarity. The longitudinal component
of the field is within
  300~G. The chemical composition and evolutionary status were studied
in the work~\cite{25:Romanyuk_n_en_4}. The lines are very narrow and the accuracy the
measurements are high.

\subsubsection {HD\,138777}

A new magnetic star with a strong field. One 2010 measurement
gave a value of +2100~G. The lines are narrow, there are many of them.
The accuracy of measurements
is high. According to the catalog~\cite{10:Romanyuk_n_en_4}, the type of peculiarity is SrEu.
There are no data on the radial velocity in the literature. Our one measurement:
$V_r=-46$~km\,s$^{-1}$.

\subsubsection {HD\,158450}

The star with a very strong field discovered by us~\cite{7:Romanyuk_n_en_4}. Four
2010 measurements showed that the extreme value of the longitudinal
component of the field reaches $-4500$~G. The magnetic monitoring of
the star was continued. The SIMBAD database shows the radial velocity
of  $V_r = -22.0$~km\,s$^{-1}$, possibly variable. In 2009, we
determined the radial velocity of $-19.3$~km\,s$^{-1}$, a mean over
2010 is  \mbox {$V_r = -16.8$}~km\,s$^{-1}$.

\subsubsection {HD\,168856}

A peculiar star with strengthened silicon lines. Its magnetic field was
found by Hubrig~\cite{26:Romanyuk_n_en_4}. The longitudinal component $B_e =-600$~G.
The details can be found in the work~\cite{3:Romanyuk_n_en_4}. In 2010, two measurements
were taken,
one of them confirmed the presence of a strong field in the star. Radial
 velocity  \mbox{$V_r=-9.8$~km\,s$^{-1}$}, given in the SIMBAD database,
differs significantly from our value of
\mbox{$V_r=-21.8$~km\,s$^{-1}$}.

\subsubsection {HD\,178892}

The magnetic field of the star was found by us in 2003, a detailed
article was published~\cite{27:Romanyuk_n_en_4}. Since the star has the strongest
field, an interest exists in regular monitoring for the purpose of
searching for long-term  $B_e$ variations. Completed in 2010, two
measurements show a strong field. No variability of radial velocity
throughout 2009 and 2010~wad  detected. The SIMBAD database does not
list any $V_r$ value.

\subsubsection {HD\,199180}

The star with the silicon and chromium peculiarities. In the 2010 observations
some  signs of the field are visible. Earlier (2009 results) we
discovered the field of the star. The radial velocity has the same value as
and a year earlier,  $V_r=-16.9$~km\,s$^{-1}$.

\subsubsection {HD\,201174}

We started the magnetic monitoring of the star in 2006 \cite{1:Romanyuk_n_en_4}. In
2010, five observations were done. The longitudinal field has a
positive sign and varies from 500~G to 2~kG. A fairly wide scatter of
radial velocities  is observed.

\subsubsection {HD\,225627}

A peculiar star with a large depression and strontium anomalies.
Earlier (in 2009) we found the star to be magnetic. In 2010,   one
  measurement out of two confirmed the presence of a field in the star.
  The lines in
the spectrum are narrow.   The radial velocity variability is within
errors.

\subsubsection {HD\,258686}

We detected the magnetic field~\cite{7:Romanyuk_n_en_4}. The field in the star is very
strong and  reaches +7~kG (longitudinal component). In 2010
one measurement is done. The  silicon peculiarity type. The rotation period
  is not found yet. ADS\,5139A is a companion at a distance of
$1\farcs 5$.
 It is a rare case when the field from   the hydrogen line core is
stronger than the one determined by the classical method. A
preliminary curve of  $B_e$ variability with a period of 1.5115~days is built.

\subsubsection {HD\,261937}

We first measured the magnetic field of the star in 2008
(the results are published in~\cite{2:Romanyuk_n_en_4}). Two 2010 measurements give
indications that the star is magnetic. For a spectral star of
class O7 it is a fairly rare case.
No data about it is listed in the catalog~\cite{10:Romanyuk_n_en_4}. A binary star.
The SIMBAD database lists it as a young
stellar object. A member of the scattered cluster NGC\,2264 aged
log t=7.0. Interstellar linear polarization of 0.2\% is noted.

\subsubsection {HD\,279021}

The magnetic star we found~\cite{2:Romanyuk_n_en_4}. The plentiful lines are narrow.
The accuracy of field measurements is high. The period is $P=2\fd 80$~days. Two
2010 measurements confirm the presence of the field. The star has a
SrCrEu peculiarity type. No data on the radial velocity  is
found in the literature.

\subsubsection {HD\,343872}

We discovered the star to be magnetic. The details are presented in
the paper with the results of 2008 observations~\cite{2:Romanyuk_n_en_4}. The field
was first found in Elkin et al.~\cite{28:Romanyuk_n_en_4}. Monitoring of the star
continues with the goal of obtaining a magnetic curve and
constructing a field model. Six measurements were made in 2008: the
field strength found by the regression method
 is on the average   15\% smaller than the one
obtained by the classical method.

\subsubsection {${\rm BD}\,+53\degr 1183$}

From the three 2010 measurements, we discovered a new magnetic star.
The lines are narrow, the period is unknown, but it exceeds several days.
Catalog~\cite{10:Romanyuk_n_en_4}  lists it as having  strontium and
chromium peculiarities. The star has been poorly studied:
there is no information in the literature about radial
velocity and rotation parameters. Judging by our data, radial
velocity
is variable, $v_e \sin i=50$~km\,s$^{-1}$.

\subsubsection {${\rm BD}\,+00\degr 4535$}

A magnetic field of 3~kG was detected by us in 2009.
Further, the presence of a strong field was confirmed: two
2010 measurements gave a field of $-0.5$~kG and 2~kG by the classical
technique and by
20\% smaller than when using the regression method. Monitoring
continues to obtain a phase magnetic curve.

\subsection{CP Stars without a Field Found in 2010 Observations}

\subsubsection {HD\,653}

A chemically peculiar star of spectral class A0. A CrEu peculiarity
type. Included in the SuperWASP list~\cite{29:Romanyuk_n_en_4}. In
this paper, a period of rotation of \mbox{$P = 1\fd 0854$}~days is
found. The first magnetic measurements were performed by us in
2008~\cite{2:Romanyuk_n_en_4}. They did not show the presence of a
field, just like the two measurements of 2010. There are numerous
lines in the spectrum, rather narrow and sharp. We found that the
star is a spectroscopic binary.
 The lines of the second component are clearly visible in
the observations made on JD = 2455488.445; in two months
the separation of components was no longer visible.
The radial velocity of the main
component in these two dates is very different.
The projection of the
 rotation velocity of the main component $v_e \sin i$ is estimated
at approximately 75~km\,s$^{-1}$. It can be assumed that the
secondary component is a slow rotator  ($v_e \sin i~\approx
20$~km\,s$^{-1}$), it is fainter by approximately one  magnitude.
According to GAIA, the parallax is  $\pi = 2.93$~mas.

\subsubsection  {HD\,23924}

A member of the Melotte\,22 cluster (the Pleiades), included into
the~\cite{10:Romanyuk_n_en_4} catalog, however, not indicating the peculiarity type.
The spectrum of A7p. The magnetic field was not
found, just like in 2007 and 2009~\cite{1:Romanyuk_n_en_4,3:Romanyuk_n_en_4}. Radial
velocity is variable, the star is a spectroscopic binary. Apparently, this
is an Am star. The further observations of the object are inexpedient.

\subsubsection {HD\,23964}

Another member of the Melotte\,22 cluster (the Pleiades). The data on it is
missing in the Renson and
Manfroid~\cite{10:Romanyuk_n_en_4}. A spectroscopic binary. No magnetic field
was found.

\subsubsection {HD\,32549}

A chemically peculiar star with   silicon and chromium anomalies. There exists
a depression at 5200~\AA. The 2010 measurements  do not reveal the presence of a
field, as well as all our previous measurements (for details, see
papers~\cite{2:Romanyuk_n_en_4,3:Romanyuk_n_en_4}).  Auriere et al.~\cite{20:Romanyuk_n_en_4} performed
a rather long series of observations of this star with a high
spectral resolution, and in some cases the authors succeeded to
register a longitudinal field of magnitude \mbox{150--200}~G.
Perhaps the object has a weak magnetic field. It seems that the radial
velocity is variable: \mbox{$V_r = +16.2$}~km\,s$^{-1}$ in the
SIMBAD; while we have obtained +31~km\,s$^{-1}$ in 2009 and two more
different values in 2010.

\subsubsection {HD\,34307}

A binary star ADS\,3857AB. It is missing in the catalog of Renson and
Manfroid~\cite{10:Romanyuk_n_en_4}. Spectral class B9. In the observations of 2010
no magnetic field was detected.

\subsubsection {HD\,34968}

A binary star. Missing in the catalog of Renson and Manfroid~\cite{10:Romanyuk_n_en_4}. A
fast rotator. In the 2010 observations no magnetic field
was detected.

\subsubsection  {HD\,35101}

The star is missing in the Renson and Manfroid catalog~\cite{10:Romanyuk_n_en_4}. No magnetic
 field was   detected in two 2010 measurements, just like in
2008~\cite{2:Romanyuk_n_en_4}. The lines in the spectrum are very wide, measurements
are difficult. The upper limit of the field is  500~G.

\subsubsection {HD\,35548}

In the Renson and Manfroid catalog~\cite{10:Romanyuk_n_en_4}  it is listed as
a~\mbox{HgMn-star}. For such objects      the lack of magnetic field
is assumed. Our 2010 observations have not revealed  the presence of
the field either. A member of the Orion\,OB1 association. A
spectroscopic binary star.

\subsubsection {HD\,35575}

The star with weakened lines of helium,a member of the Orion\,OB1a association.
A fast rotator, the   line  profiles  are complex, there are few of them.
Accurate field measurements
  are difficult. In 2010, one measurement was carried out, which
 did not show the presence of a magnetic field. Our measurement of radial
velocity, $V_r = +24$~km\,s$^{-1}$  significantly differs from the
value given in the SIMBAD database \mbox{($V_r = +9$~km\,s$^{-1}$)}.

\subsubsection {HD\,35730}

The star with weakened lines of helium, a member of the Orion\,OB1a association.
The spectrum has a few fairly narrow, strong, symmetrical
lines. The measurements of 2010   have not   revealed the field.

\subsubsection {HD\,36032}

It is missing in the catalog~\cite{10:Romanyuk_n_en_4}.  Apparently, we observe a normal
class B9 star.
The reason for the star being in the list of observations is unclear.
The star spatially falls into the  Orion\,OB1 association, however,
there is no evidence of its membership in it. An attempt to measure the field
in the classical way revealed that this can only be done
from two broad lines with complex profiles.
We failed to find the field from a single spectrum.

\subsubsection {HD\,36629}

A magnetic star; the field was first discovered by Borra~\cite{11:Romanyuk_n_en_4}.
Spectral class B3, a He-wk peculiarity~\cite{19:Romanyuk_n_en_4}. The spectrum
reveals very narrow and sharp lines. The star is immersed in the
Parenago\,1044 nebula, in whose direction a strong interstellar (or
circumstellar) linear polarization  is observed \mbox{($P=
1.843$\%)}. According to the 2010 measurement, we have not detected
any  field. The longitudinal field was either not discovered later
(see Fig.~\cite{30:Romanyuk_n_en_4}).

\subsubsection {HD\,36982}

A very hot star, a member of the youngest subgroup d of the
Orion\,OB1 association. The age of the star is less than 1~Myr. The
spectrum reveals a strong linear polarization
  $P = 1.007$\%. We failed to find the magnetic field. Presumably the
  longitudinal field
of about 100~G was found at the FORS1
VLT~\cite{18:Romanyuk_n_en_4}.The SIMBAD database lists the radial velocity  of \mbox
{$V_r = +38$}~km\,s$^{-1}$, which differs significantly from
the one found by us \mbox {$V_r=+12.6$}~km\,s$^{-1}$.

\subsubsection {HD\,37525}

The star with weakened lines of helium, a member of the Orion\,OB1b
association. The brightest star in the  $\sigma$\,Orion cluster. The
age of this cluster is estimated at 1.2~Myr. In 2010 it was observed
once and  magnetic field was not found. The lines are very wide ($v_e
\sin i = 150$~km\,s$^{-1}$). The measurements by the classical method
are impossible to carry out, the regression method has neither
yielded any field detection. In the SIMBAD database, the star is
represented as a young stellar object of spectral class B5. A
companion is found at a distance of $0\farcs 5$.

\subsubsection  {HD\,38104}

 A chemically peculiar star with CrEu anomalies. It is described in detail in
our work~\cite{2:Romanyuk_n_en_4}. Since 2005, we have carried out eight observations,
no magnetic field was detected. The lines are very narrow,
the measurement accuracy is high,    hence, based on
our measurements, we estimate the upper limit of
the field  to be 300~G. Apparently, the radial velocity
is variable, which suggests the binarity of the star.

\subsubsection {HD\,39317}

Two measurements did not yield the presence of a magnetic field in this
SiEuCr star. The results of previous years are published in
Romanyuk et al.~\cite{2:Romanyuk_n_en_4}. Over five years the star does not show
the presence of a longitudinal field. It is unlikely that this is a very slow
rotator, the magnetic field of the star  is  most likely below our
detection limit. The radial velocity found in our work,  \mbox
{$V_r = -12.1$}~km\,s$^{-1}$  differs from the one given in the SIMBAD database,
$V_r = -4.4$~km\,s$^{-1}$.

\subsubsection {HD\,52711}

The magnetic field was not detected in the observations, just as before.
The details can be found in Romanyuk et al.~\cite{2:Romanyuk_n_en_4}.

\subsubsection {HD\,62512}

Two 2010 measurements did not show the presence of a field, just like
earlier~\cite{2:Romanyuk_n_en_4}. In the SIMBAD database $V_r=-4$~km\,s$^{-1}$. It means
that the radial velocity is variable. Apparently, this is a binary
star.

\subsubsection {HD\,90763}

A chemically peculiar star with strengthened Sr lines. Three
measurements performed for three consecutive nights showed that
the upper limit of the longitudinal field is $B_e = 300$~G. $V_r =
-24.8$~km\,s$^{-1}$ in the SIMBAD database. Our three values are close to the
specified value. It is now too early to estimate the variability of   radial
velocity.

\subsubsection {HD\,93294}

A poorly studied binary peculiar star. It is not represented in the Renson and
Manfroid~\cite{10:Romanyuk_n_en_4}. Two 2010 measurements
show that if the magnetic field of the star exists,
its longitudinal component does not exceed 300~G. The star is binary.

\subsubsection {HD\,97633}

HD\,97633 =  $\theta$\,Leo. Measurements in December 2010 (three nights in a
row) showed no field. The upper limit does not exceed
200~G. It does not make sense to examine the star for a magnetic
field any longer.

\subsubsection {HD\,108506}

A cool chemically peculiar star. Observations for three
consecutive nights in December 2010 did not lead to the discovery of the field:
the upper limit is~1~kG. The lines are wide. In the SIMBAD database it is listed
as the $\delta$\,Sct type.
In the~\cite{10:Romanyuk_n_en_4} catalog, the spectral
class F1 and the SrCr type of peculiarity are marked.
The radial velocity  $V_r =
-5.4$~km\,s$^{-1}$ (SIMBAD) does not on average differ from our
2010 measurements, but there is reason to consider it variable.
The projection of the rotation velocity on the line of sight  $v_e \sin i =
150$~km\,s$^{-1}$ is similar to the one given in the VIZIER database.

\subsubsection {HD\,114125}

An  Algol-type eclipsing binary. In the~\cite{10:Romanyuk_n_en_4} catalog,
spectral class F2 and the SrEuCr type of peculiarity are listed.
Two measurements of the
field  completed in 2010 yielded  a zero result.
The upper limit of the field is~200~G. The SIMBAD lists \mbox {$V_r =
-33$}~km\,s$^{-1}$, which is significantly different from the value we obtained.

\subsubsection {HD\,135297}

A chemically peculiar star with sharp lines.
Very strong anomalies of strontium and chromium are noted in the spectrum.
The photometric index is  \mbox {$\Delta a  = 0.032$}. This
   is characteristic of magnetic stars. Two of our 2010 measurements
 gave a zero result. The SIMBAD gives
\mbox {$V_r =
-31.9$}~km\,s$^{-1}$, which  significantly differs from the
 measurement results we have obtained. It is possible that the star is binary.

\subsubsection {HD\,149046}

The star is poorly studied. In the Renson and Manfroid catalog~\cite{10:Romanyuk_n_en_4}
it is designated as SrCrEu-peculiar. One observation of 2010 showed
the absence of a field. The radial velocity determined by us \linebreak
\mbox {$V_r=-28$}~km\,s$^{-1}$  is  close to the one given in the SIMBAD database
($V_r = -23.5$~km\,s$^{-1}$).

\subsubsection {HD\,157740}

A bright  CrEuSr-type star. For the search for a magnetic field was not earlier
done. Our one measurement in December 2010 gave
a zero result. The radial velocity is close to that given in the
SIMBAD database  $V_r=+11.2$~km\,s$^{-1}$.

\subsubsection {HD\,178308}

A peculiar star with anomalous chromium lines. Two
2010 measurements did not lead to the discovery of the field.
The SIMBAD database   gives the value
\mbox {$V_r=-30.4$}~km\,s$^{-1}$; our two measurements are in
agreement with each other, but differ from the SIMBAD data, which can
testify about the binarity of the star.

\subsubsection {HD\,182180}

A very fast rotator, $v_e \sin i=300$~km\,s$^{-1}$. One
2010 measurement provides evidence of the existence of a magnetic field.
Observations should be continued. Broad lines do not allow to
perform accurate measurements.

\subsubsection {HD\,198920}

The star with peculiarities of strontium and europium. Just like the year
earlier (2009), in the reporting year no magnetic field was detected.

\subsubsection {HD\,217401}

A binary system ADS\,16437AB. Peculiarity in Sr. We have previously found it to be
 magnetic  (2009 results) with a longitudinal field of about
$-400$~G. In 2010 the field was not found. Radial velocity is
variable.

\subsubsection {HD\,281367}

A peculiar SrEu star. The first observations were made in 2008, they
did not reveal the presence of the field. One measurement of 2010 also did
not bring any evidence of the existence of the field. A fast rotator.

\subsubsection {${\rm BD}\,+38\degr 2360$}

A cool peculiar SrCrEu-star. Studied poorly. The lines are narrow.
The magnetic field  was not detected in 2010.

\subsubsection {${\rm BD}\,+37\degr 431$}
A spectroscopic  binary. An F2 spectrum. One measurement of 2010 did not reveal the
presence of the field. The accuracy of the measurements is high.

\subsubsection {${\rm BD}\,+36\degr 363$}

A member of the NGC\,752 (RV\,14) cluster. The spectral class of the star is F2p,
peculiarity of Sr. One 2010 measurement    did not lead to the field detection.

\subsubsection {${\rm BD}\,-12\degr 2366$}

A member of the NGC\,2539 cluster. It was  detected as magnetic as a
result of our 2008 observations.  Two 2010 measurements did not show
the presence of the field.

\subsubsection {NGC\,752-105}

The star from the NGC\,752 cluster. It is not included in the catalog
of Renson and Manfroid~\cite{10:Romanyuk_n_en_4}. The field was not found in 2010,
the lines in the spectrum are wide.

\section {CONCLUSION}

In 2010 we have therefore performed observations of 92 stars with
a Zeeman analyzer: four of them are non-magnetic
standard stars, five are  standard magnetic stars with well-studied
longitudinal field curves. In 47 stars the 2010 observations
revealed the presence of a field, and in other 36 it was not detected. we
discovered twelve new magnetic stars, and in other three stars the
field is  suspected.

In 2010, a large program for the study of magnetic stars in the
Orion\,OB1 cluster was launched. Polarized spectra of over 20 objects
of the association were obtained.

Most of the observed stars are poorly studied, therefore
we measured radial velocities and projections
of  rotation velocities to the line of sight for all objects.
A comparison with the literary
data revealed a number of new binary stars.

\begin{acknowledgements}
The authors are grateful to G.~A.~Chuntonov for his assistance in
preparation for observations and their provision. The authors thank
Russian Science Foundation for the financial support of the study
 (an RSF grant no.~14-50-00043).
\end{acknowledgements}


\end{document}